ORIGINAL ARTICLE

# Enriching physical-virtual interaction in AR gaming by tracking identical objects via an egocentric partial observation frame

**Liuchuan Yu[1] · Ching-I Huang[2] · Hsueh-Cheng Wang[3] · Lap-Fai Yu[1]**

**Abstract**

Augmented reality (AR) games, particularly those designed for head-mounted displays, have grown increasingly prevalent. However, most existing systems depend on pre-scanned, static environments and rely heavily on continuous tracking or marker-based solutions, which limit adaptability in dynamic physical spaces. This is particularly problematic for AR headsets and glasses, which typically follow the user's head movement and cannot maintain a fixed, stationary view of the scene. Moreover, continuous scene observation is neither power-efficient nor practical for wearable devices, given their limited battery and processing capabilities. A persistent challenge arises when multiple identical objects are present in the environment–standard object tracking pipelines often fail to maintain consistent identities without uninterrupted observation or external sensors. These limitations hinder fluid physical-virtual interactions, especially in dynamic or occluded scenes where continuous tracking is infeasible. To address this, we introduce a novel optimization-based framework for re-identifying identical objects in AR scenes using only one partial egocentric observation frame captured by a headset. We formulate the problem as a label assignment task solved via integer programming, augmented with a Voronoi diagram-based pruning strategy to improve computational efficiency. This method reduces computation time by 50% while preserving 91% accuracy in simulated experiments. Moreover, we evaluated our approach in quantitative synthetic and quantitative real-world experiments. We also conducted three qualitative real-world experiments to demonstrate the practical utility and generalizability for enabling dynamic, markerless object interaction in AR environments. Our video demo is available at https://youtu.be/RwptEfLtW1U.

## 1 Introduction

Augmented reality (AR) integrates 3D virtual objects into the real world in real time (Azuma 1997). AR games have become increasingly prevalent in recent years as AR hardware and software advance. AR games can generally be divided into two categories according to the devices they run on: mobile AR games and headset AR games. Mobile AR games are designed to be played on mobile devices such as smartphones and tablets. They use the on-device camera to scan the real world and the screen to overlay digital content onto the real world, enhancing the user's experience. A notable example of mobile AR games is Pokémon GO. Headset AR games require a headset to provide an immersive AR experience. The headset typically has a display that visualizes digital content within the user's field of vision, allowing for interactive and engaging gameplay. Figure 1 shows some examples.

✉ Liuchuan Yu
lyu20@gmu.edu

Ching-I Huang
cihuang@gapp.nthu.edu.tw

Hsueh-Cheng Wang
hchengwang@g2.nctu.edu.tw

Lap-Fai Yu
craigyu@gmu.edu

[1] Department of Computer Science, George Mason University, 4400 University Drive, Fairfax, VA 22030, USA

[2] Power Mechanical Engineering Department, National Tsing Hua University, 101, Section 2, Kuang-Fu Road, Hsinchu 300044, Taiwan

[3] Electrical and Computer Engineering Department and Institute of Electrical and Control Engineering, National Yang Ming Chiao Tung University, No. 1001, Daxue Rd. East Dist., Hsinchu 300093, Taiwan

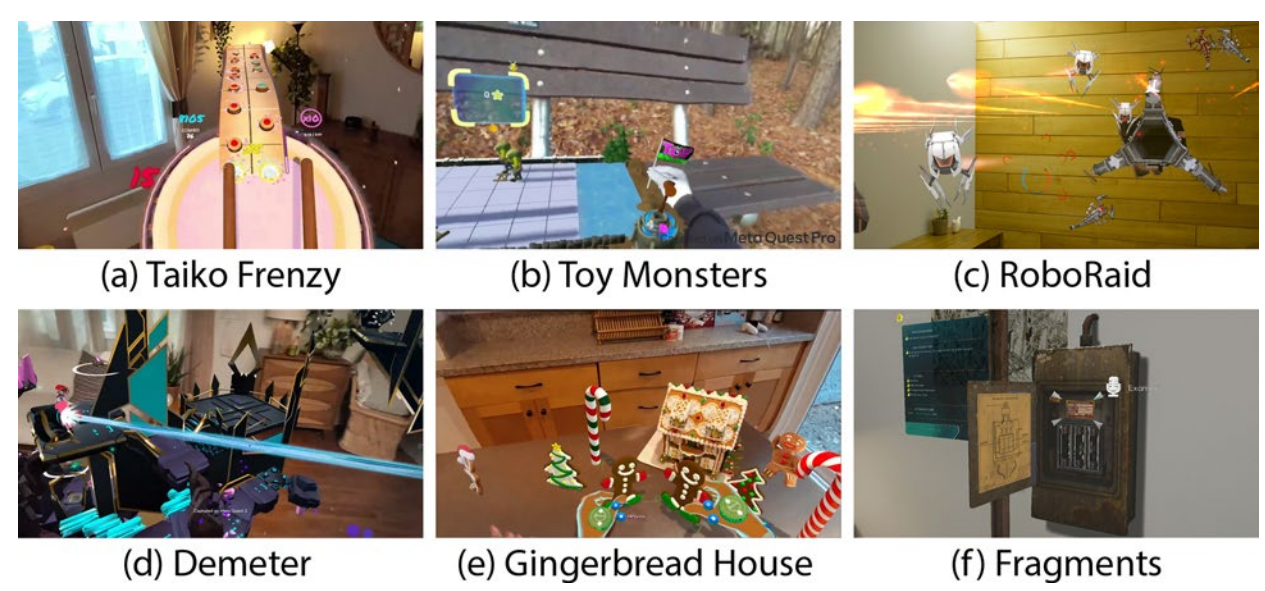

**Fig. 1** Example headset AR games. Image courtesy of the game developers

AR games have found applications in various domains, including education (Pellas et al. 2019; Nincarean et al. 2013; Weerasinghe et al. 2019; Amanatidis 2022), accessible learning (Cai et al. 2021), rehabilitation and exercise (Nelson and Gabbard 2023), medicine and healthcare (Ducasse et al. 2019), and digital art (Lichty 2019). Further insights can be gleaned from comprehensive reviews on the subject (Geroimenko 2019; Marto and Gonçalves 2022). Traditionally, AR games follow the practice of scanning the scene first and subsequently overlaying virtual objects. However, the real scene layout is susceptible to intentional or unintentional manipulation. Additionally, prevalent interaction methods in AR games involve controllers or hand-tracking, with limited exploration of physical-virtual interaction. While a virtual proxy is feasible, as highlighted in Simeone et al. (2015), the disparity between real and virtual objects can impact user experience, a critical aspect in gaming. Therefore, there is a compelling need to support dynamic physical environments and enable seamless physical-virtual interactions without requiring re-scanning or continuous scene observation. This motivates our investigation into efficient identity tracking for physical objects that may move or be rearranged during gameplay.

On the other hand, identical objects are ubiquitous, both in the physical and virtual realms. Whether they are identical tables and chairs in offices, classrooms, or restaurants, or identical items in real toy sets, or even in games where numerous non-player characters (NPCs) and props share identical characteristics, tracking the identities of these objects presents a practical challenge. Feature-based multiple object tracking (MOT) often falls short of reliably distinguishing object identities when the objects share identical appearances or characteristics. Moreover, traditional MOT approaches typically rely on stationary cameras and/or continuous observation, rendering them unsuitable for AR headset-based applications. Specifically, AR headsets and glasses are egocentric in nature, meaning the camera constantly moves with the user and cannot maintain a fixed viewpoint of the scene. Furthermore, continuous visual observation is not only computationally demanding but also power-inefficient, making it impractical for mobile or wearable devices with limited energy and processing capacity.

Although some marker-based AR games offer a solution for tracking identical objects, they require the cumbersome process of printing and affixing markers, which is not ideal for widespread AR applications. Furthermore, considering the importance of portability and energy efficiency, it might be impractical to introduce additional devices that continuously observe the scene, potentially hindering the gaming experience. These constraints highlight the need for a lightweight, markerless, and observation-efficient tracking method that works reliably in dynamic environments. Hence, it is worth exploring a cost- and power-efficient approach to track identical objects seamlessly without relying on additional devices.

To address these challenges, we propose a lightweight optimization-based approach capable of distinguishing visually identical object instances within a dynamic scene. Our method leverages the assumption that, in the context of AR games, most objects maintain relatively stable spatial relationships (i.e., relative positions and orientations) with each other over time, enabling robust object differentiation solely based on spatial relationship reasoning. Figure 2 illustrates our approach with an example.

Rather than relying on fixed cameras capturing global scene changes, our approach exclusively leverages partial observations from the user's AR headset. The egocentric frame serves as an input to our integer programming-based formulation, which reassigns object identities by minimizing a cost function rooted in spatial consistency. Such observations enable our integer programming-based approach to resolve object identities and update object poses, ensuring an accurate understanding of the evolving scene layout. Notably, our approach operates without the need for continuous observation. Furthermore, our approach incorporates the Voronoi diagram to expedite computations, enhancing scalability for large game scenes. Quantitative experiments both in synthetic environments and real-world environments were conducted to validate the effectiveness of our approach. We demonstrate the practical utility of our method through three real-world qualitative experiments, including an AR farm-to-table game, an AR storytelling scenario, and a robot delivery scenario, to illustrate its generalizability.

Our contributions are as follows:

- Proposing a lightweight optimization-based framework for re-identifying visually identical objects in AR scenes using an egocentric partial observation frame, without requiring continuous tracking or external devices.
- Formulating the problem as an integer programming task based on spatial consistency and introducing a

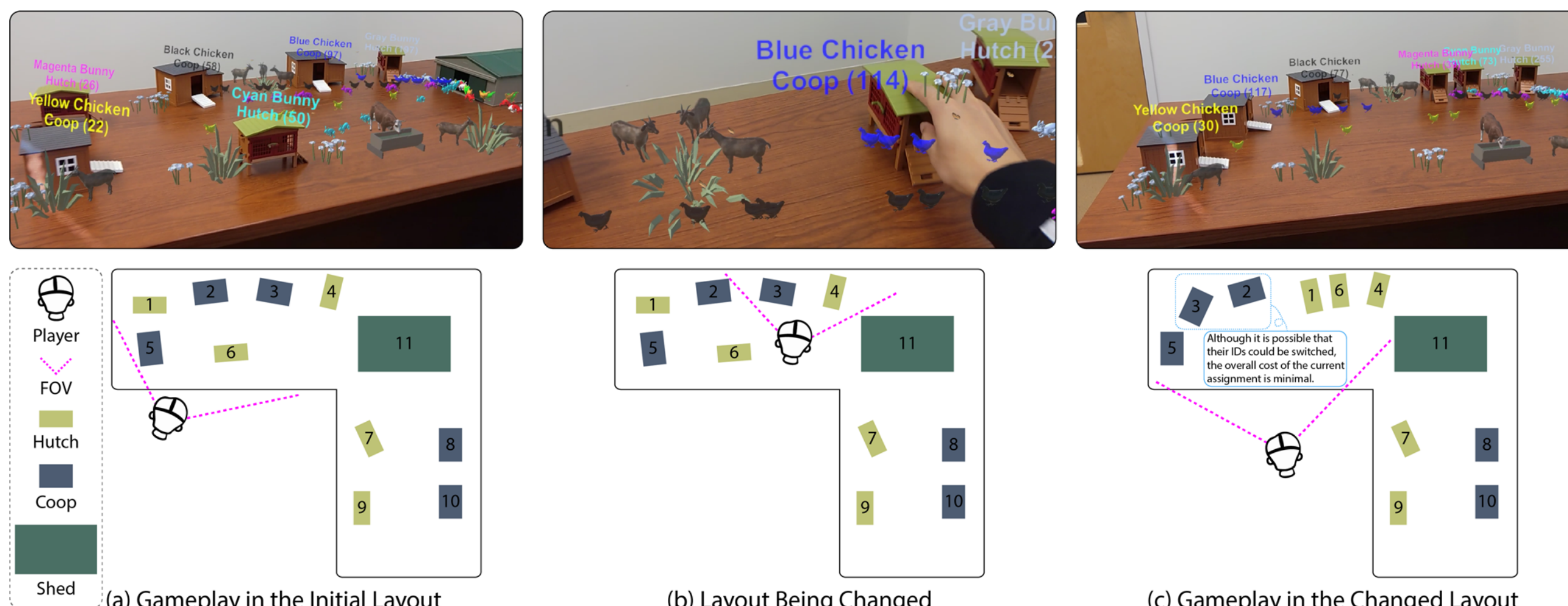

**Fig. 2** Our approach empowers AR games to adapt to dynamic scenes with changing real object layouts. It is based on the assumption that, in the context of AR games, most objects maintain relatively stable spatial relationships (i.e., positions and orientations) with each other. In this illustration, we begin by scanning the layout and creating a farm-to-table AR game. This scene encompasses multiple identical objects, such as hutches and coops. The first row presents the AR view, whereas the second row illustrates the corresponding spatial layout representations with labeled object IDs. The gameplay in the initial layout, depicted in (**a**), features labels in the format of *animal building (count)*, where *animal building* indicates the type of colored animals attracted to the building and *count* denotes the number of such colored animals that have entered the building. Coops and hutches cater to chickens and bunnies, respectively, and animals move toward their designated buildings. For instance, the label *Yellow Chicken Coop (22)* in **a** signifies that this coop (ID:5) is for yellow chickens, and 22 yellow chickens have entered the coop. As shown in (**b**), the real objects' layout is being manipulated by the player. **c** Our approach keeps track of the identities of the buildings so that the AR game continues to function seamlessly after the layout change. For example, the blue chicken coop (ID:3) from (**a**) is still recognized after it is moved to the left as shown in (**c**), even though it looks identical to the other coops. Note that it is possible that some IDs (e.g., ID:3 and ID:2) could be switched, but the overall cost of the current assignment is minimal and hence selected

Voronoi diagram-based pruning strategy to improve computational efficiency.

- Validating our approach through quantitative synthetic and real-world experiments and a farm-to-table AR game, demonstrating its accuracy, efficiency, and applicability to various AR scenarios such as storytelling and robot delivery.

## 2 Related work

Augmented reality serves as an effective tool to blend physical and virtual worlds. By overlaying 3D models, animations, and texts that align with a user's physical surroundings, AR provides an immersive experience. Interactions with physical objects using AR predominantly involve object detection and 6DOF object tracking. These methodologies are crafted to identify specific objects and track their positions and orientations. We discuss the related techniques in this section.

### 2.1 Object detection for AR content authoring

A fundamental challenge of augmented reality pertains to showing realistic content around the user. This challenge, in the realm of computer vision, largely relies on semantic segmentation and object detection (Lee et al. 2021) for preprocessing. For an immersive AR experience that provides real-time augmentation, such preprocessing algorithms need to run fast enough to catch up with the moving speed of humans (Ko and Lee 2020). Conventional methods that apply feature tracking algorithms such as SIFT, or pixel-by-pixel classification of handcrafted features, such as support vector machines, faced challenges in achieving optimal segmentation performance (Lee et al. 2021). Conversely, neural networks, particularly CNNs such as FasterRCNN (Ren et al. 2015), the YOLO series (Redmon and Farhadi 2018; Simony et al. 2018), and SSD detectors (Liu et al. 2016), have witnessed substantial progress and are now seamlessly integrated into AR applications (Zhang et al. 2020; Tanzi et al. 2021). They are proficient in handling occlusion challenges within the AR domain such as overlaying virtual objects (Roxas et al. 2018).

Furthermore, AR applications such as gaming and storytelling (Liang et al. 2021; Li et al. 2022, 2023; Li and Yu 2023) that employ virtual characters need to place the characters with respect to the semantics and poses of objects in the scene to deliver realistic experiences. To ensure logical interaction with objects, considering object poses is critical. Object orientations might be deduced from the bounding boxes of scene objects integrated within a volumetric map derived from RGB-D streams (Tahara et al. 2020).

However, object detection and pose estimation methods typically focus on class-level recognition and are not inherently designed to resolve instance-level identities among multiple identical objects. This constrains its suitability for more expansive AR applications.

## 2.2 Physical-virtual interaction

Due to the nature that AR overlays virtual objects onto the physical environment, the interaction between the physical environment and virtual objects has attracted increasing research interests. Similar to VR, most headset-based AR applications, including games, use controllers, eye-tracking, haptics, voice control, or hand-tracking to enable interaction with virtual objects (Balakrishnan et al. 2021). AR games that involve real object interactions are relatively less explored but promise to gain traction.

There are research efforts in the physical-virtual interaction direction. For example, Simeone et al. (2015) investigated substitutional reality where virtual counterparts substitute the physical world and found that mismatches between virtual and physical objects would become an obstacle to the interaction and user experiences. Lee et al. (2019) designed an actuator system underneath the table to synchronize physical and virtual objects in a face-to-face mixed reality environment. Hu et al. (2020) proposed a prototype to let users directly interact with physical objects, which provided active reactions in AR interactions. Hartmann et al. (2019) presented RealityCheck to blend the player's real-world surroundings with the virtual world. Kaimoto et al. (2022) proposed Sketched Reality to support bi-directional interactions between AR-based virtual sketches and actuated tangible UIs. Li et al. (2019) presented IDCam to fuse radio frequency identification (RFID) and computer vision to bridge objects in the physical world with digital information in AR. Jain et al. (2023) presented UbiTOUCH, an AR system designed to use physical objects as tangible proxies for virtual objects. This system focuses on hand-object interaction, analyzing how users naturally interact with objects to recommend suitable proxies that provide haptic feedback and maintain consistency between physical and virtual interactions.

However, prior works require devices in addition to the headset to enable physical-virtual interaction. Some devices are customized, such as in Lee et al. (2019); Hu et al. (2020). Other devices are commercially available, such as in Hartmann et al. (2019); Kaimoto et al. (2022); Li et al. (2019). Besides, some suffer from visual distractions and extreme mapping cases (Jain et al. 2023).

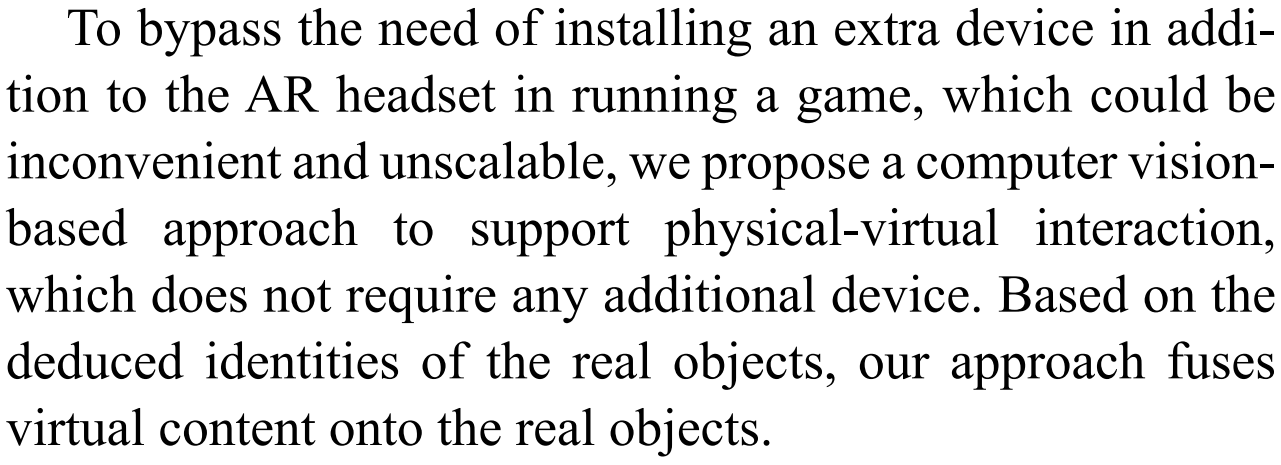

To bypass the need of installing an extra device in addition to the AR headset in running a game, which could be inconvenient and unscalable, we propose a computer vision-based approach to support physical-virtual interaction, which does not require any additional device. Based on the deduced identities of the real objects, our approach fuses virtual content onto the real objects.

## 2.3 Object tracking in augmented reality

Object tracking is a major research area in computer vision, with multiple object tracking (MOT) representing a specialized subdomain focused on concurrently tracking multiple objects (Park et al. 2021). Traditional MOT methods rely on continuous observations of captured video sequences and often employ learning-based approaches to achieve accurate tracking, such as (Meinhardt et al. 2022; Ren et al. 2023). However, these conventional techniques are not well-suited for tracking identical objects in AR gaming environments due to several limitations: (1) continuous observation is impractical in dynamic AR settings; (2) the AR camera is not stationary and moves with the player, resulting in partial and non-static observations; and (3) MOT methods based on video sequence analysis can be computationally expensive, making them challenging for real-time gaming applications that demand low-latency performance.

Object tracking in AR encompasses both marker-based and markerless methods. Fiducial marker tracking, a fundamental approach, relies on artificial markers affixed to an object's surface for user-friendly tracking purposes (Kato and Billinghurst 1999). Despite the widespread use of marker-based methods, such as QR code-based tracking, it is generally impractical to put markers on every object in a scene. Depending on the device used for tracking, object tracking for AR can be categorized into mobile AR tracking and headset AR tracking. While numerous works focus on mobile AR tracking, less research has been conducted on AR tracking methods that use headsets alone.

In the realm of mobile AR, various approaches have been proposed. Mooser et al. (2007) introduced an efficient and accurate object-tracking algorithm based on graph cut segmentation, eliminating the need for a preexisting 3D model. Park et al. (2008) presented a method to simultaneously track multiple 3D objects by combining object detection and tracking. Rambach et al. (2017) introduced the concept of Augmented Things, where objects carry necessary tracking and augmentation information. Le et al. (2021) incorporated machine learning for detecting and tracking AR marker targets. Lee et al. (2022) proposed a system enabling camera tracking in the real world, visualizing virtual information through object recognition and positioning. Arifitama et al. (2021) investigated markerless-based tracking as a potential substitute for marker-based tracking in AR problems. For a comprehensive overview of mobile AR tracking, please refer to the review by Syed et al. (2022). On the other hand, research on AR tracking methods based on headsets alone is relatively scarce. Frantz et al. (2018) explored the augmentation of HoloLens with the Vuforia image processing SDK for neuronavigational use. Radkowski (2018) integrated the HoloLens into a point cloud-based tracking system using Kinect range cameras.

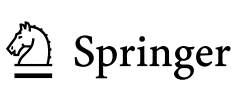

These AR tracking methods typically focus on locating objects or enabling interaction but do not address the persistent identification of multiple identical objects, while real environments such as offices, apartments, and classrooms commonly consist of multiple identical objects (e.g., chairs, desks). Given the prevalence of AR headsets, it is important to investigate tracking methods, particularly markerless tracking methods, that use AR headsets only and are capable of deducing the object identities of multiple visually-identical objects in an environment.

## 3 Technical approach

Figure 3 shows the pipeline of our technical approach. In the beginning, we partition the scene via a Voronoi diagram, which is regarded as the search space for possible labels of detected objects in the changed layout. In the initial layout, the Voronoi partitioning and IDs are manually assigned. The Voronoi sites are defined by semantic clusters. As objects are moved in the scene, an AR camera captures images of the scene based on which objects are detected. The pose estimation algorithm computes the objects' 6DOF poses in the camera coordinate system. Since the pose of the AR camera in the world coordinate system is known, the detected objects can be transformed from the camera coordinate system to the world coordinate system. After that, our approach calculates three cost terms: translation cost, rotation cost, and dimension cost between possible objects in the initial layout (in the world coordinate system) and detected objects in the changed layout (also in the world coordinate system). These costs are the input of the integer programming algorithm, which determines the identities of the detected objects.

### 3.1 6DOF pose estimation

The prerequisite for accurately assigning labels is estimating the 6DOF poses of objects. Previous studies have relied on either RGB, RGB-D, or point cloud data (Liu et al. 2024). Notable off-the-shelf implementations, such as Objectron (Ahmadyan et al. 2021), are available. If there is no off-the-shelf pose estimation solution, we have devised a computer vision-based pipeline as shown in Fig. 22.

### 3.2 Object label assignment algorithm

*Voronoi Diagram Partitioning* Considering the efficiency and spatial structure of the environment, we utilize the Voronoi diagram to split the whole space into several partitions. Let $S_k$ be the $k$-th site in the Voronoi diagram, $C_k$ be the center point of the $k$-th site, and $P$ be the AR camera's position. Let $D_p^k$ be the distance between the camera's position and the center point of the $k$-th site, we have:

$$D_p^k = |P - C_k|. \tag{1}$$

According to the Voronoi diagram's properties, if the camera is within a site $S_k$, it means that among all site centers,

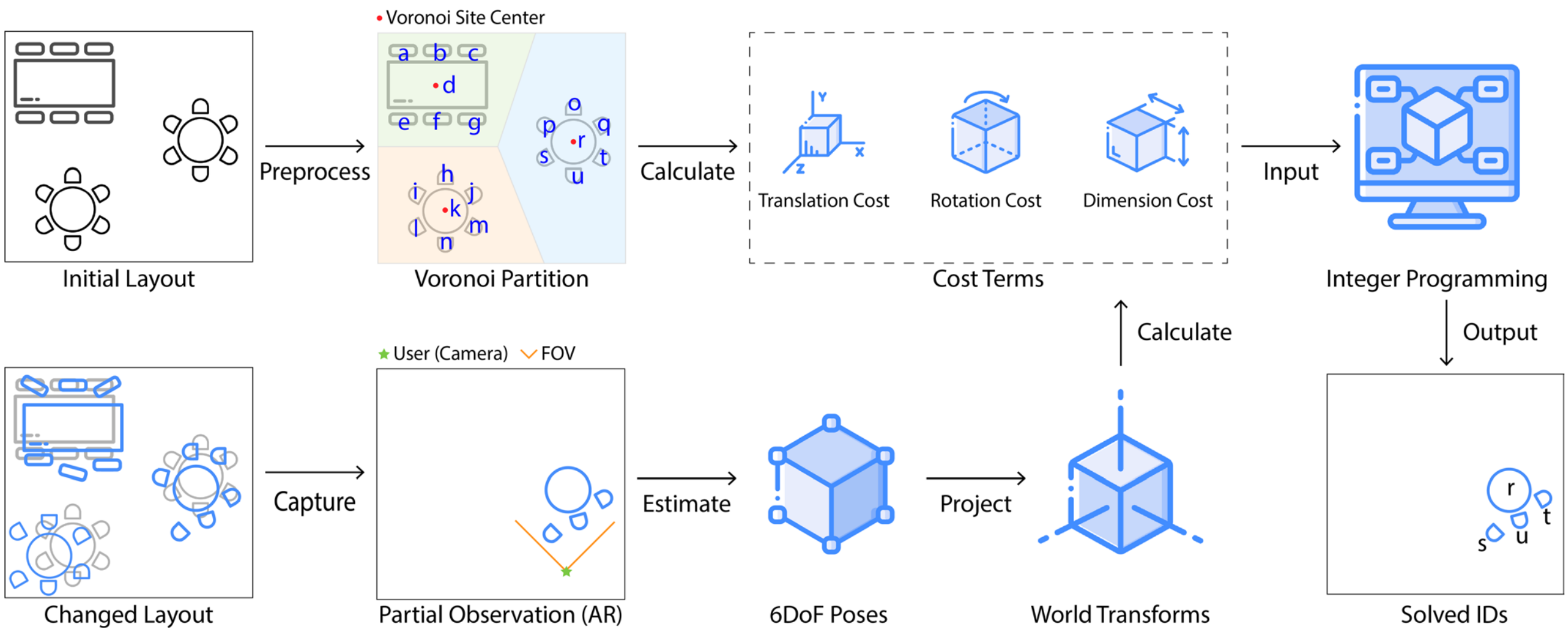

**Fig. 3** The pipeline of our approach. Please refer to the main text for explanation

site $S_k$'s center $C_k$ is the nearest to the camera. We define an indicator function *within* as:

$$within(P, S_k) = \begin{cases} 1 & \text{if } D_p^k \text{if is the minimum among all sites,} \\ 0 & \text{otherwise.} \end{cases} \quad (2)$$

Based on the current AR camera's position, we define the probability of considering site $S_k$ as:

$$Prob(S_k) = \begin{cases} 1 & \text{if } within(P, S_k) = 1, \\ \frac{\frac{1}{D_p^k}}{\sum_t \frac{1}{D_p^t}} & \text{if } within(P, S_k) = 0. \end{cases} \quad (3)$$

In the following, **possible objects** refer to the objects in the initial layout, whereas **detected objects** denote those identified by the camera in the changed layout. This implies that detected objects are within the camera's field of view (FOV).

*Search Space Pruning* An intuitive approach to solving the label assignment problem is to assign each detected object a label selected from all possible objects. However, considering the fact that the number of detected objects is less than the number of possible labels, it would be inefficient to search through all labels and assign them to detected objects. The camera's position is helpful in pruning the possible label space since the detected results of objects farther away from the camera tend to be less accurate. Therefore, we leverage the Voronoi diagram partitioning result to help filter the possible labels, as follows: First, we sort the sites' probabilities in descending order. Second, we apply a threshold to filter out sites that are far away from the camera. Third, we only use those objects within the filtered result for cost computation and integer programming.

As we formulate the object tracking problem as a label assignment problem solved by integer programming, we define the translation, rotation, and dimension costs for label assignment evaluations.

*Translation Cost* For simplicity, we assume that objects are only moved horizontally, not vertically. For example, a chair is pushed around the ground plane but not lifted. In other words, objects are moved in the x and z directions but not the y direction. We define the translation cost as:

$$C_\mathrm{t} = \frac{\sqrt{(x - \hat{x})^2 + (z - \hat{z})^2}}{l}, \quad (4)$$

where $(x, z)$ and $(\hat{x}, \hat{z})$ denote the positions of the possible object and the detected object, respectively on the xz-plane. $l$ denotes the diagonal length of the scene, which is used for normalizing the cost.

*Rotation Cost* We assume that an object is only rotated about its $y$-axis. Accordingly, the rotation cost is defined as:

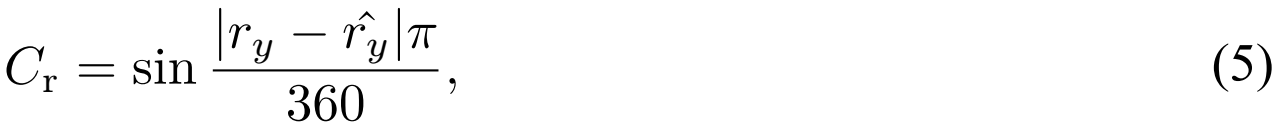

$$C_\mathrm{r} = \sin\frac{|r_y - \hat{r_y}|\pi}{360}, \quad (5)$$

where $r_y$ and $\hat{r_y}$ represent the rotation (in degree) about the $y$-axis of the possible object and the detected object, respectively. We use the sine function to normalize the cost to [0, 1] because (1) the maximal rotation cost happens when the object is rotated by 180° and (2) the rotation cost is symmetrical, meaning that it should have the same cost if the object is rotated by 90° or 270° (equivalent to rotating by 90° in the counter-clockwise direction).

*Dimension Cost* We assume that the detected objects may belong to different categories (e.g., chairs, desks) with different dimensions. So we define a cost term named dimension cost $C_\mathrm{d}$. We use $w$ for width, $h$ for height, and $d$ for depth to describe the dimension information of the bounding box of a detected 3D object. Then, we define the dimension cost $C_\mathrm{d}$ as follows:

$$C_\mathrm{d} = \frac{\max(w, \hat{w})}{\min(w, \hat{w})} \times \frac{\max(h, \hat{h})}{\min(h, \hat{h})} \times \frac{\max(d, \hat{d})}{\min(d, \hat{d})}, \quad (6)$$

where $w$ denotes the width of the possible object and $\hat{w}$ denotes the width of the detected object. Similar for $h$ and $d$, which refer to the height and depth.

In Eq. (6), $C_\mathrm{d}$ is always greater than or equal to 1. Because of the possible rotation of the 3 axes in the pose estimation results (represented as bounding boxes), the width, height, and depth of possible objects and detected objects could be misaligned. Therefore, we use the minimal value to represent this cost.

*Integer Programming* We formulate this object label assignment problem as an integer programming problem. Each detected object should be assigned a proper label, and each label should be used at most once. In other words, one label cannot be assigned to more than one detected object.

Let $N$ be the total number of all detected objects and $M$ be the total number of possible labels. Let $i \rightarrow j$ refer to assigning the $i$-th detected object (within the field of view of the camera) with the $j$-th possible label (within all possible Voronoi sites).

Let $C_\mathrm{t}$ be the translation cost; $C_\mathrm{r}$ be the rotation cost; and $C_\mathrm{d}$ be the dimension cost. Let $w_\mathrm{t}$ be the weight of the translation cost and $w_\mathrm{r}$ be the weight of the rotation cost. Then, we use $C^{i \rightarrow j}$ to denote the total cost of assigning the $i$-th detected object with the $j$-th possible label:

$$C^{i \rightarrow j} = C_\mathrm{d}^{i \rightarrow j}(w_\mathrm{t} C_\mathrm{t}^{i \rightarrow j} + w_\mathrm{r} C_\mathrm{r}^{i \rightarrow j}). \quad (7)$$

The objective function refers to minimizing the total cost of assigning a label to every detected object:

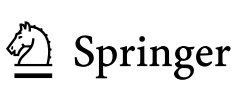

$$\min \sum_{i=1}^{N}\sum_{j=1}^{M} C^{i\to j}A^{i\to j},$$
$$s.t. \sum_{j=1}^{M} A^{i\to j} = 1, \forall i \in N, \tag{8}$$
$$\sum_{i=1}^{N} -A^{i\to j} >= -1, \forall j \in M.$$

Here, $A^{i\to j}$ denotes whether the assignment exists or not:

$$A^{i\to j} = \begin{cases} 1 & \text{if } i \to j \text{if exists,} \\ 0 & \text{otherwise.} \end{cases} \tag{9}$$

## 4 Experiments

To assess the effectiveness of our approach, we undertake a comprehensive examination through both quantitative and qualitative experiments. Our quantitative experiments involve the simulations of scenes with varying complexity levels and randomized translation/rotation manipulation using Gaussian noise models. Specifically, we utilize office and restaurant layouts, which are typically characterized by a multitude of identical objects such as tables and chairs. In the qualitative experiments, we implemented a Farm-to-Table AR game. Furthermore, we showcased the potential applications in AR storytelling and robot delivery, providing a broader perspective on the general applicability of our approach.

The code of our approach is available at https://github.com/gmudcxr/EnrichingARInteraction. We also provide a video that shows our experiment results at https://youtu.be/RwptEfLtW1U.

### 4.1 Quantitative synthetic experiments

*Implementation* The system prototype for synthetic quantitative experiments was developed on Windows 11 and Unity 2020.3.20 using an Alienware Aurora R12 with 32GB RAM, Intel i7-11700F CPU, and Nvidia RTX 3070. For integer programming, we use lp_solver[1] as the solver.

**Table 1** A summary of the six synthetic scenes

| Complexity Level | Scene Name | No. Sites | No. Object Types | No. Objects | Scene Size ($m^2$) |
|---|---|---|---|---|---|
| Low | L1 | 2 | 1 | 16 | 25 |
| Low | L2 | 4 | 1 | 8 | 64 |
| Medium | M1 | 5 | 3 | 35 | 64 |
| Medium | M2 | 6 | 2 | 30 | 25 |
| High | H1 | 10 | 5 | 52 | 49 |
| High | H2 | 14 | 11 | 48 | 225 |

#### 4.1.1 Different scenes

*Layout Types* We simulate two types of layouts, restaurants and offices, which are typically full of identical objects. Based on the number of sites, the number of object types, the number of objects, and the scene size, we create 3 different complexity levels: (L)ow, (M)edium, and H(igh). In total, 6 scenes were used in our experiments. Table 1 shows their statistics. Figure 4 shows their top-down views.

*Movement Simulation* We use the Unity asset, Bézier Path Creator, to create the wandering path and set a speed to simulate the movement of the AR camera.

*Metric* To evaluate the performance of our approach, we calculate the accuracy as a metric:

$$accuracy = \frac{N_c}{N_d}, \tag{10}$$

where $N_c$ denotes the number of correct labels and $N_d$ denotes the number of detected objects.

*Adding Noises* If the initial scene layout does not change, the assignment will always succeed by returning the initial label assignment as the solution, which corresponds to a cost of zero. To simulate object manipulations in the layout, we apply Gaussian noise to every object.

We assume that one object can be moved by translation and rotation, corresponding to the aforementioned translation cost and rotation cost. For translation, we assume that

[1] lp_solve: https://lpsolve.sourceforge.net/5.5/

★ Start/End Point — Wandering Path ← Moving Direction

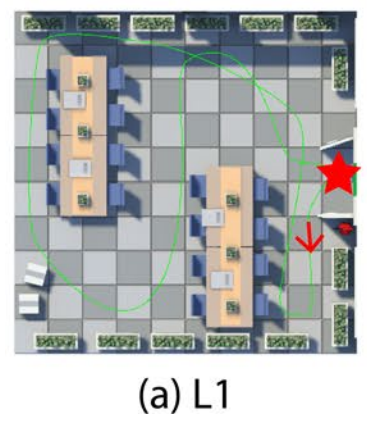
(a) L1

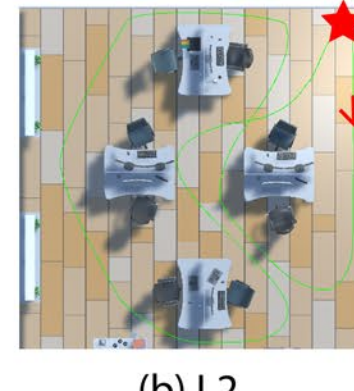
(b) L2

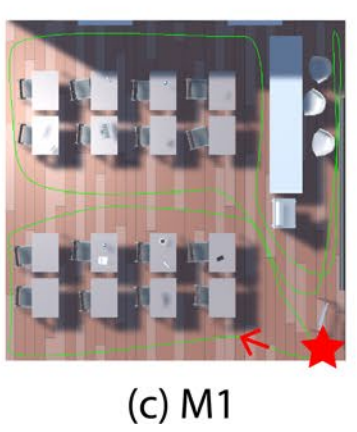
(c) M1

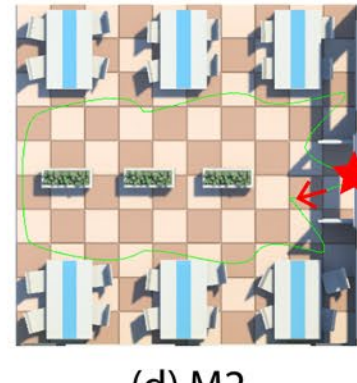
(d) M2

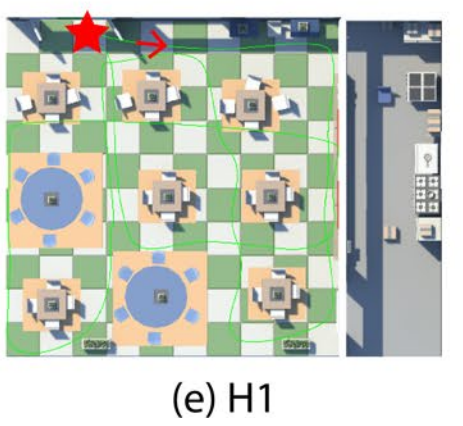
(e) H1

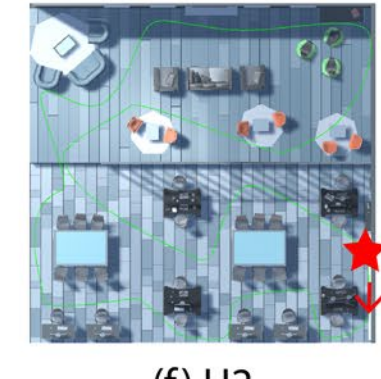
(f) H2

**Fig. 4** Screenshots of the six scenes with different complexity levels. Note that although shown in approximately the same size for visual clarity, the layouts are scaled down by different amounts due to different original sizes

the object can move along the *x* and *z* directions. Since we conduct simulations with Unity, we use the Unity coordinate system here, where the *y*-axis points up. As for rotation, we assume that an object only rotates about the *y*-axis.

We use $T(t_{\mathrm{m}}, t_{\mathrm{s}})$ to denote the translation noise model, where $(t_{\mathrm{m}}, t_{\mathrm{s}})$ stands for the mean and standard deviation (SD) of the Gaussian noise model for translation *x* and *z*. Although *x* and *z* share the same noise model parameters, they are independent models and work independently. We use $R(r_{\mathrm{m}}, r_{\mathrm{s}})$ to denote the rotation noise model with a mean of $r_{\mathrm{m}}$ and a standard deviation of $r_{\mathrm{s}}$. We add constraints to the translation noises and rotation noises to mimic realistic object movements. If the position of an object after applying translation noise is out of the room boundary, we will force resetting its position to be within the boundary. As for the rotation noise for object rotation, we restrict it to be within $[-360^{\circ}, 360^{\circ}]$.

*Experiment Procedure* For each scene, we execute the following procedure. **(a)** Pick a translation SD value *a* from $TList = \{0.1, 0.2, 0.3, ..., 1, 2, 3, ..., \sqrt{scene_size}\}$ to fill the translation noise model *T*(0, *a*); **(b)** Pick a rotation SD value *b* from $RList = \{0, 5, 10, ..., 120\}$ to fill the rotation noise model *R*(0, *b*), using $360/3 = 120$ as the maximum value of rotation model SD as data falling outside of three SDs is rare; **(c)** Use *T*(0, *a*) and *R*(0, *b*) to generate a noisy layout; **(d)** Let the camera move along the pre-defined path with a constant speed and objects' identities will be calculated every 100 frames; **(e)** Reset the layout once the camera reaches the end; **(f)** Loop the previous steps until all values from *TList* and *RList* have been used once. The translation cost weight $w_{\mathrm{t}}$ and the rotation cost weight $w_{\mathrm{r}}$ were set to $0.36\sqrt{scene_size}$ and 1, respectively.

*Result Analysis* Fig. 5 shows the results. Overall, the accuracy drops as the SD of the translation noise model or rotation noise model increases, which is reasonable because the SD value increases the messiness of the layout.

We observe that, even using the same translation noise model, the accuracies of the same complexity level vary. For example, as shown in Fig. 5a1 and b1, the accuracy curves in (b1) are above those in (a1). An obvious difference between scene L1 (Fig. 4a) and scene L2 b is that chairs in L1 are more clustered than those in L2.

On the other hand, a more complex scene does not necessarily imply a lower accuracy compared to a less complex scene. For example, medium-complexity scene M1 (Fig. 4c) has a better accuracy than low-complexity scene L1 (Fig. 4a) at the low translation noise level. The reason is that objects in M1 are sparser than those in L1. A similar trend is observed when comparing high-complexity scene H2 with low-complexity scene L1 as shown in Fig. 5f1 and a1.

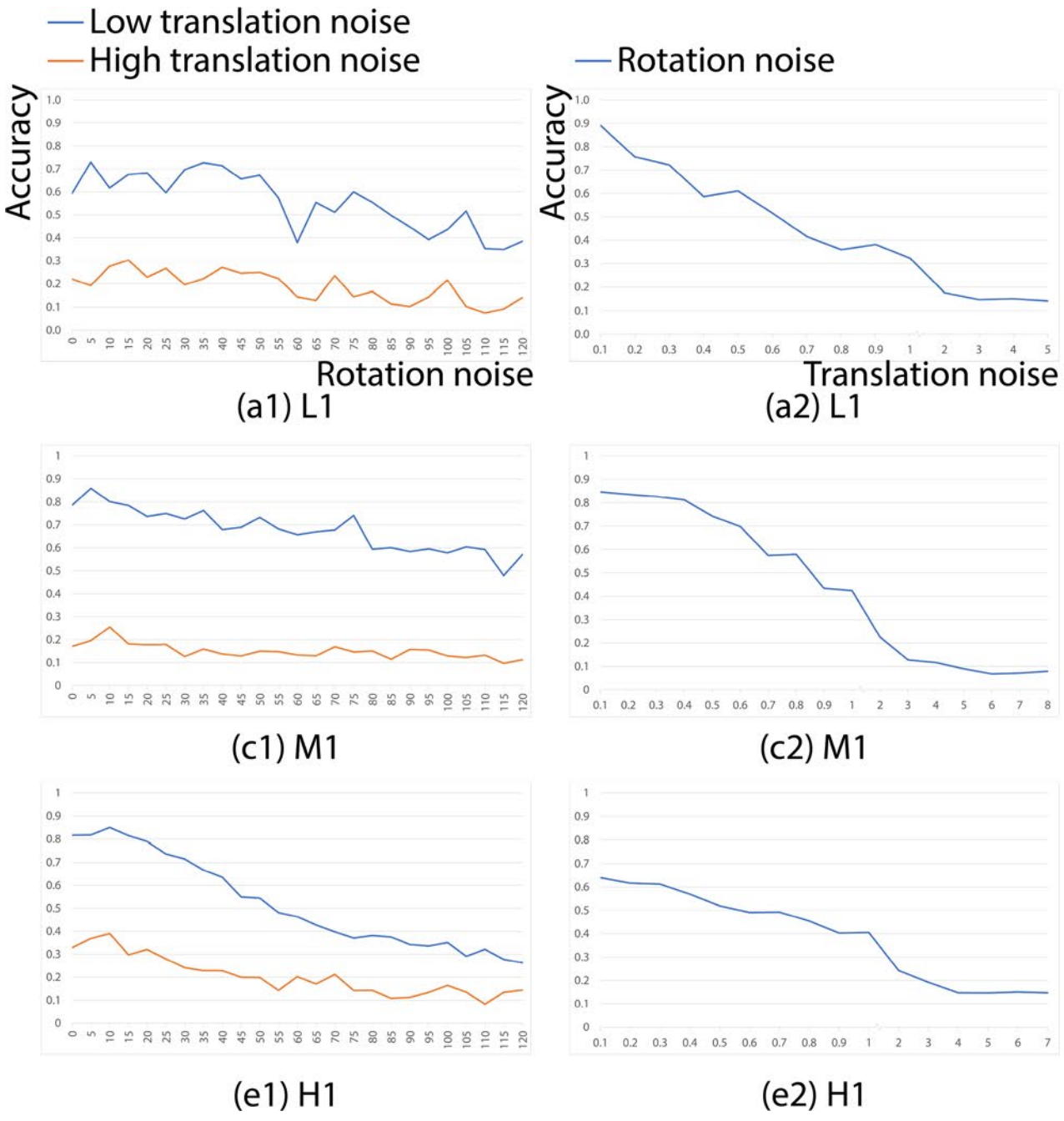

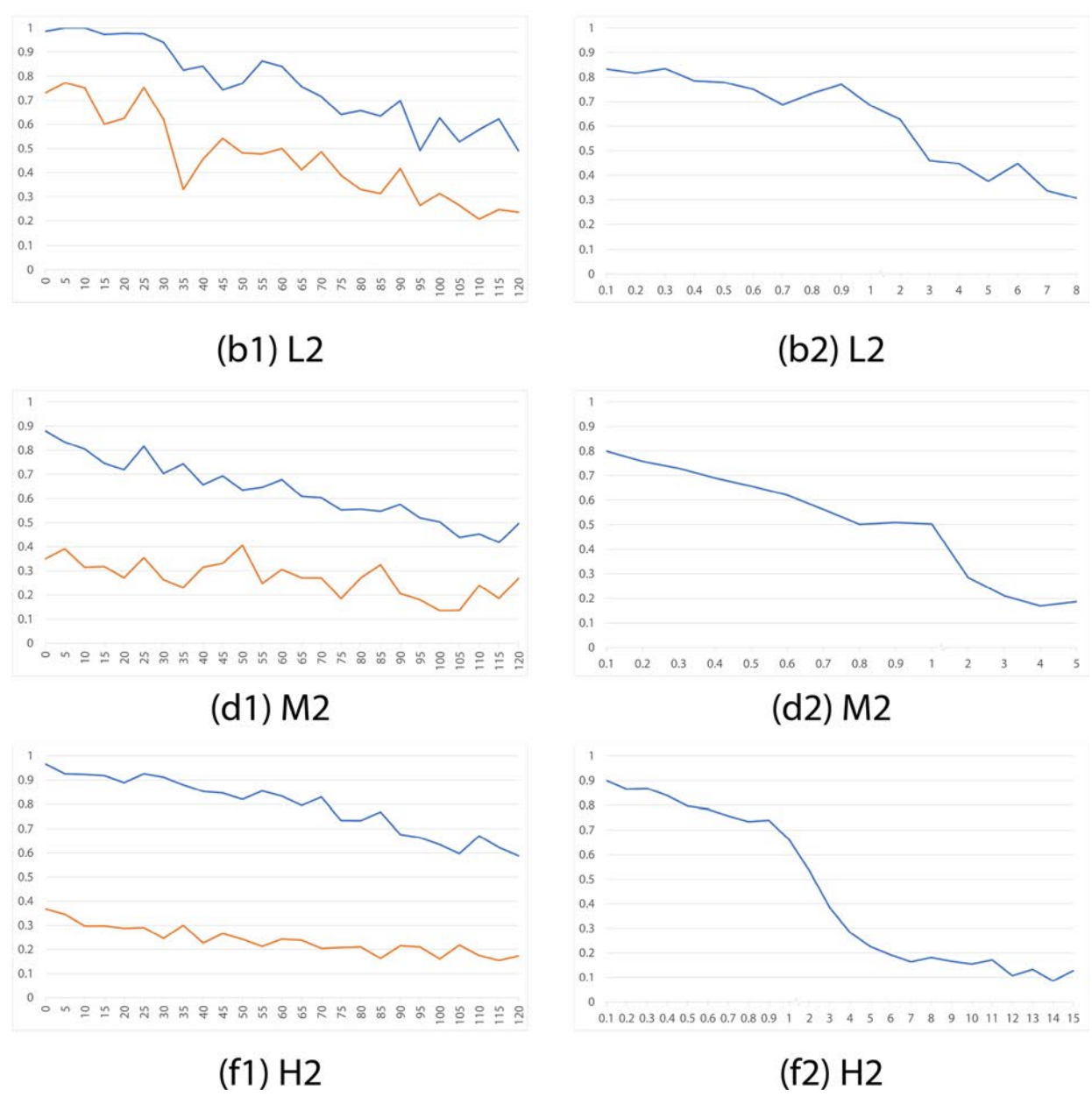

**Fig. 5** Accuracies of the six scenes with different noise levels. Each scene is associated with two figures, e.g., low-complexity-level scene L1 has two figures, (a1) and (a2). The first figure (a1) depicts the accuracy across different rotation noise levels. Low translation noise refers to noise models from *T*(0, 0.1) to *T*(0, 1.0) with an SD increment of 0.1, while high translation noise refers to noise models from *T*(0, 1.0) to $T(0, \sqrt{scene_size})$ with an SD increment of 1.0. Each point represents the mean accuracy of different translation noise models. The second figure (a2) shows the accuracy of different translation noise levels. Each point represents a mean value of mean accuracies of rotation models from *R*(0, 0) to *R*(0, 120) with an SD increment of 5

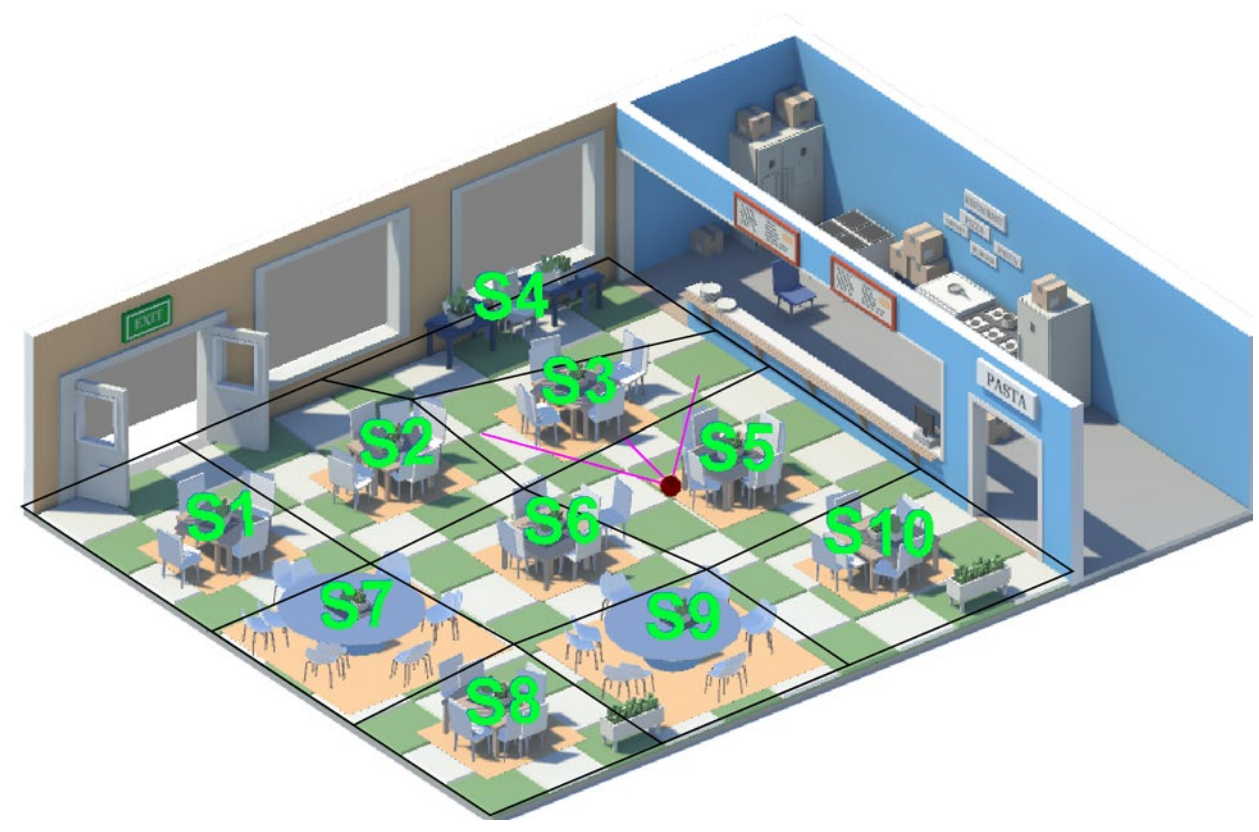

**Fig. 6** Demonstration for the ablation study and the search space pruning experiments

When the SD of translation noises increases, the results will become steady when the accuracy does not keep dropping, as shown in Fig. 5a2, c2, and e2. This is because the objects will reach the boundaries of the scene due to a large translation noise. Therefore, if a lower SD pushes objects to the boundaries, a higher SD also pushes objects to the boundaries. This means that applying the two different translation noise models makes little difference. Obviously, a low translation noise always leads to better accuracy compared to a high translation noise in the same scene.

In general, our approach has a better performance in the office scenes (L2, M1, and H2) than in the restaurant scenes (L1, M2, and H1). It makes sense because an office is usually sparser than a restaurant. For the office scenes (L2, M1, and H2), if the rotation noise SD is 50°, the average accuracy of low translation noise levels is 77.39%. If the translation noise SD is 0.5 m, the average accuracy of all rotation noise levels is 77.23%.

The results show that our approach works well at different layout complexity levels. If the layout of objects is sparse and less clustered, our approach generally performs better.

#### 4.1.2 Ablation study

To systematically evaluate the contribution of individual cost termsnamely, translation cost ($C_{\mathrm{t}}$), rotation cost ($C_{\mathrm{r}}$), and dimension cost ($C_{\mathrm{d}}$)–within our approach, we conducted an ablation study. First, we employed *T*(0, 0.3) and *R*(0, 30) to introduce noise into the layout based on scene H1 (Fig. 4e). The resulting noisy layout is illustrated in Fig. 6. Black lines mark the boundaries of the Voronoi diagram. S1 to S10 refer to the sites of the Voronoi diagram. The red dot shows the AR camera's position. The two long pink lines indicate the camera's FOV, and the short pink line indicates the camera's orientation.

**Table 2** Means and standard deviations of the accuracy under different conditions

| Condition | Mean | STD |
|---|---|---|
| w/ All Costs ($C_{\mathrm{t}}$, $C_{\mathrm{r}}$, $C_{\mathrm{d}}$) | **0.91** | **0.19** |
| w/o Translation Cost ($C_{\mathrm{t}}$) | 0.17 | 0.27 |
| w/o Rotation Cost ($C_{\mathrm{r}}$) | 0.76 | 0.27 |
| w/o Dimension Cost ($C_{\mathrm{d}}$) | 0.85 | 0.24 |

The camera was set to traverse a predefined path at a constant speed. Every 100 frames, accuracy was measured under four conditions: with all cost terms included (w/ All Costs), without the translation cost (w/o Translation Cost), without the rotation cost (w/o Rotation Cost), and without the dimension cost (w/o Dimension Cost). The camera then continued along its path, periodically recalculating accuracy every 100 frames until reaching the end of the trajectory. In total, 62 data points were collected throughout this process.

The means and standard deviations of the accuracy under different conditions are shown in Table 2. The findings indicate that the translation cost has the most significant impact on accuracy, followed by the rotation cost and, subsequently, the dimension cost.

Figure 7 presents an example from the ablation study. Figure 7a depicts the input noisy layout, while Fig. 7b shows the ground truth IDs. Figures 7c–f display the calculated IDs under four different conditions. The numbers in parentheses correspond to the accuracies under different conditions. For instance, the value 1.00 in Fig. 7c indicates an accuracy of 1.00 under the condition with all costs.

The results show that the translation cost plays a crucial role in accuracy, as its removal resulted in a 86% decrease in the example of Fig. 7. This effect is followed by a 57% accuracy reduction when the rotation cost is excluded and a 29% decrease when omitting the dimension cost. These findings are consistent with the results presented in Table 2.

Overall, the ablation study confirms that the three proposed cost terms are essential for accurately determining the correct IDs in our approach.

#### 4.1.3 Search space pruning experiments

We conduct a thresholding experiment to validate the Voronoi diagram-based search space pruning method. We use the same procedure as the aforementioned synthetic experiment except that (1) we only use noise models *T*(0, 0.1) and *R*(0, 15); (2) the wandering path is traveled once; and (3) when the camera stops, we will iterate each threshold from 0 to 1 by a step of 0.05, set the threshold, get the possible objects, and count the time cost of running the integer programming 100 times. We conducted this experiment on the H1 scene using $w_{\mathrm{t}} = 2.52$ and $w_{\mathrm{r}} = 1$.

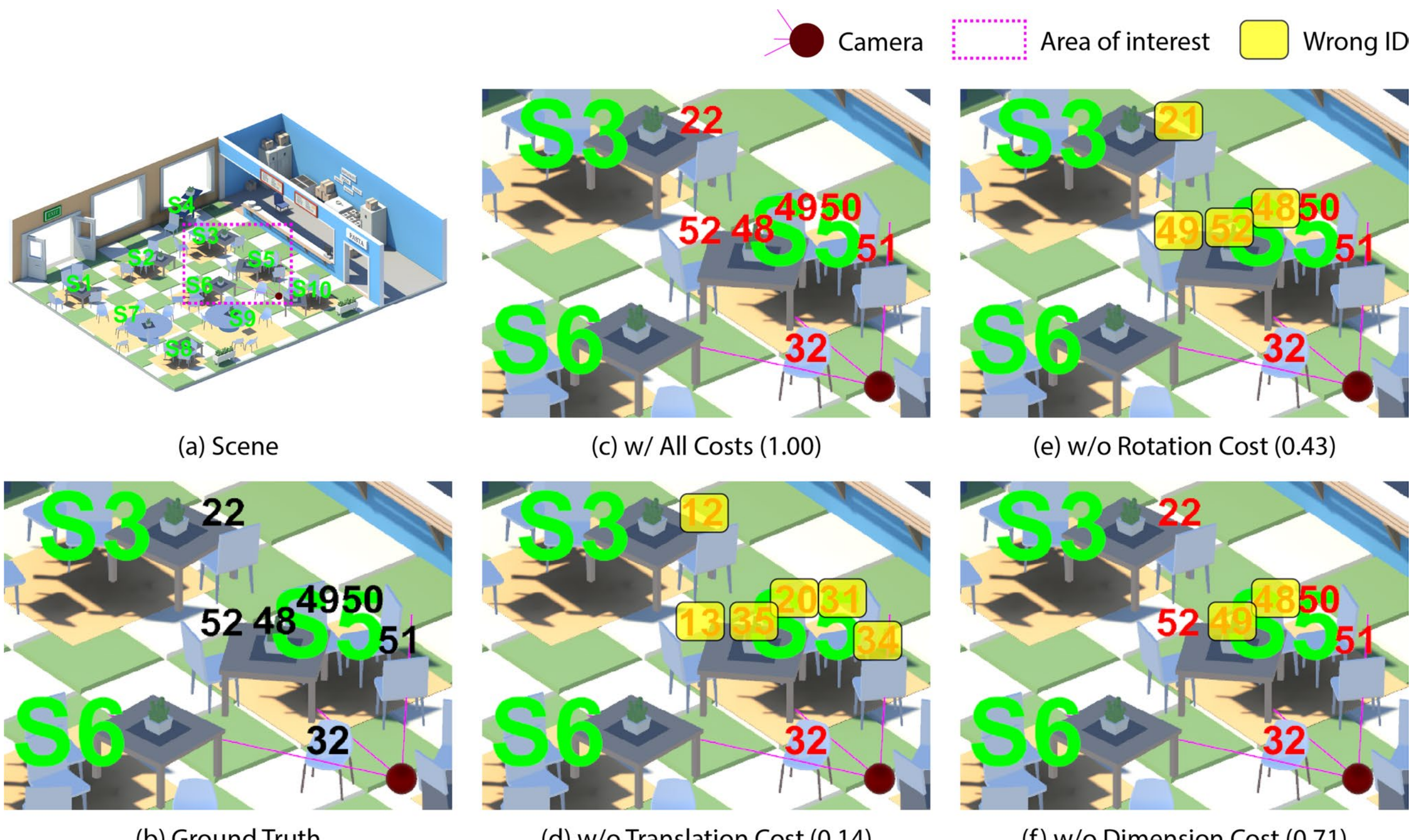

**Fig. 7** An illustration of our ablation study. The outer pink lines and the central pink line within the camera view represent the camera's FOV and orientation, respectively. The area of interest within the camera's FOV is highlighted. The noisy layout is shown in (**a**). The black numbers in (**b**) denote the ground truth IDs, whereas the red numbers in (**c**)–(**f**) indicate the assigned IDs under four different conditions. The numbers in parentheses, such as 1.00, represent the accuracies under the corresponding conditions

**Table 3** Sites' probabilities and cumulative probabilities at the moment shown in Fig. 6

| Site | S6 | S3 | S10 | S9 | S2 | S4 | S7 | S1 | S8 |
|---|---|---|---|---|---|---|---|---|---|
| Probability | 0.214 | 0.150 | 0.129 | 0.110 | 0.109 | 0.080 | 0.076 | 0.067 | 0.065 |
| Cumulative Probability | 0.214 | 0.364 | 0.494 | 0.604 | 0.713 | 0.793 | 0.868 | 0.935 | 1.000 |

Note that site S5's probability is 1 as the camera is located within site S5.

Because the thresholds are set by fixed intervals and the sites' cumulative probabilities do not fit those intervals well, we take the following measurements.

Those sites whose cumulative probability is firstly greater or equal to the threshold $t(t > 0)$ are regarded as the sites within the threshold. For example, as shown in Table 3 and Fig. 8, when $t$ equals 0, only site S5 is considered. When $t$ equals 0.50, sites S5, S6, S3, S10, and S9 are considered because the cumulative probability is 0.604, which is just greater than 0.50.

We use average accuracy and average time cost to denote the accuracy and time cost of one threshold. Figure 9 shows the result.

This experiment is intuitive. As shown in Eq. (8), whatever the threshold is changed to, the total number of all detected objects, $N$, is always the same. The difference depends on $M$, which is the total number of possible labels. When the threshold increases, more sites will be considered, leading to more possible labels. In the objective function, the cost terms increase, and so do the constraints. Therefore, the approach needs more time to solve the integer programming problem. Since the camera has a limited range, considering those objects far away from the camera does not contribute to the accuracy. That is why as the threshold reaches 0.75, the accuracy attains 1.0.

Our search space pruning method based on the Voronoi diagram works for this task. When the threshold is properly set, our approach takes less time to achieve the same performance. This trade-off between time and accuracy could facilitate applying our approach to a large scene with many objects.

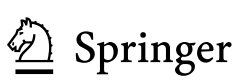

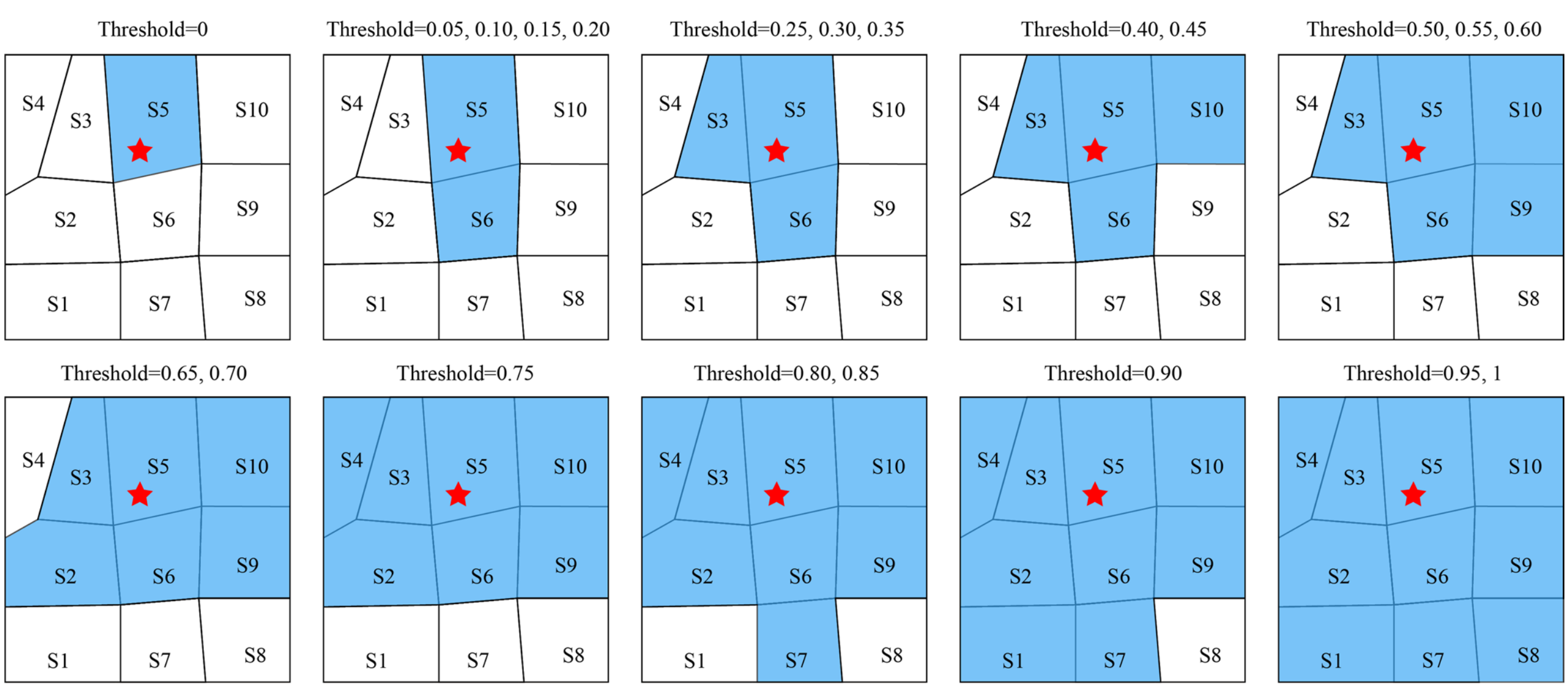

**Fig. 8** An illustration of the threshold experiment. This is plotted based on Fig. 6. S1-S10 refer to sites. The red star indicates the camera's position. The cell filled with blue color is considered under the corresponding threshold value. For example, as shown in the top-left sub-figure, only S5 is considered when the threshold is 0. As shown in the top-right sub-figure, S3, S5, S6, S9, and S10 are considered when the threshold is 0.50, 0.55, or 0.60

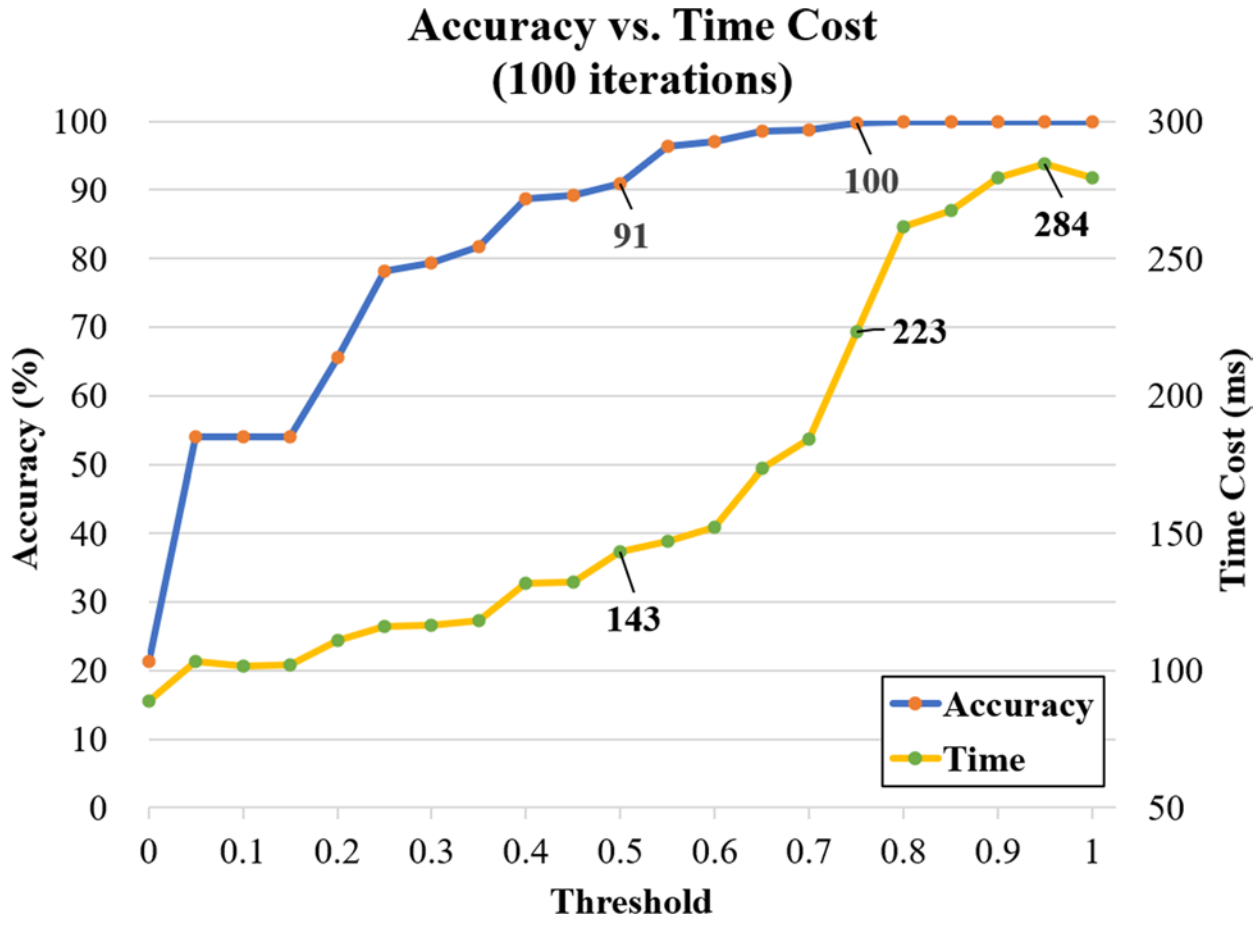

**Fig. 9** Accuracy vs time cost. When the threshold reaches 0.50, the average accuracy reaches 91%, which only takes about 50% of time compared with the no-threshold method (i.e., threshold = 1.0). When the threshold reaches 0.75, the average accuracy reaches 1.00, which saves about 20% of the time cost compared to not using a threshold for pruning

**Table 4** The three real-world scenes

| Scene | #Sites | #Chairs | Size ($m^2$) |
|---|---|---|---|
| Pantry | 2 | 10 | 18 |
| Cafeteria | 5 | 16 | 49 |
| Classroom | 7 | 29 | 130 |

## 4.2 Quantitative real-world experiments

We also conducted quantitative real-world experiments in a built environment, comprising three distinct scenes: the pantry, the cafeteria, and the classroom. The characteristics of these scenes are summarized in Table 4.

*Implementation* We used the same development environment as in the synthetic experiments. Unlike the synthetic experiments, we developed a standalone application that runs directly on the Microsoft HoloLens 2 (HL2). Log information, such as accuracy and running time, was collected and stored on the HL2 for further analysis.

*Experiment Procedure* We first scanned the environment to obtain the initial layout. Tables were designated as the Voronoi sites, and identical chairs were assigned unique integer identifiers (IDs). Figure 11 illustrates the initial layout of the pantry scene and its corresponding Voronoi partition from a top-down view. Figure 10 presents the AR view on the HoloLens 2, where $S1$ and $S2$ denote the Voronoi sites, and numbers 1–10 indicate the assigned IDs. Figure 12 shows the AR views of the cafeteria and classroom scenes on the HoloLens 2.

Chairs were then moved and rotated. The assigned IDs served as the ground truth, enabling accuracy evaluation by comparing the assigned IDs with the solved IDs. The user wore the HoloLens 2 and walked around the layout. IDs of chairs within the field of view (FOV) of the HoloLens 2 were solved in real time, with the solving process executed every frame. Both the time cost and accuracy were logged to an on-device file. Additionally, the HoloLens 2 view was recorded during application runtime. Figure 13 presents selected frames from the pantry scene with the changed layout.

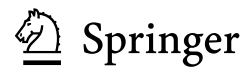

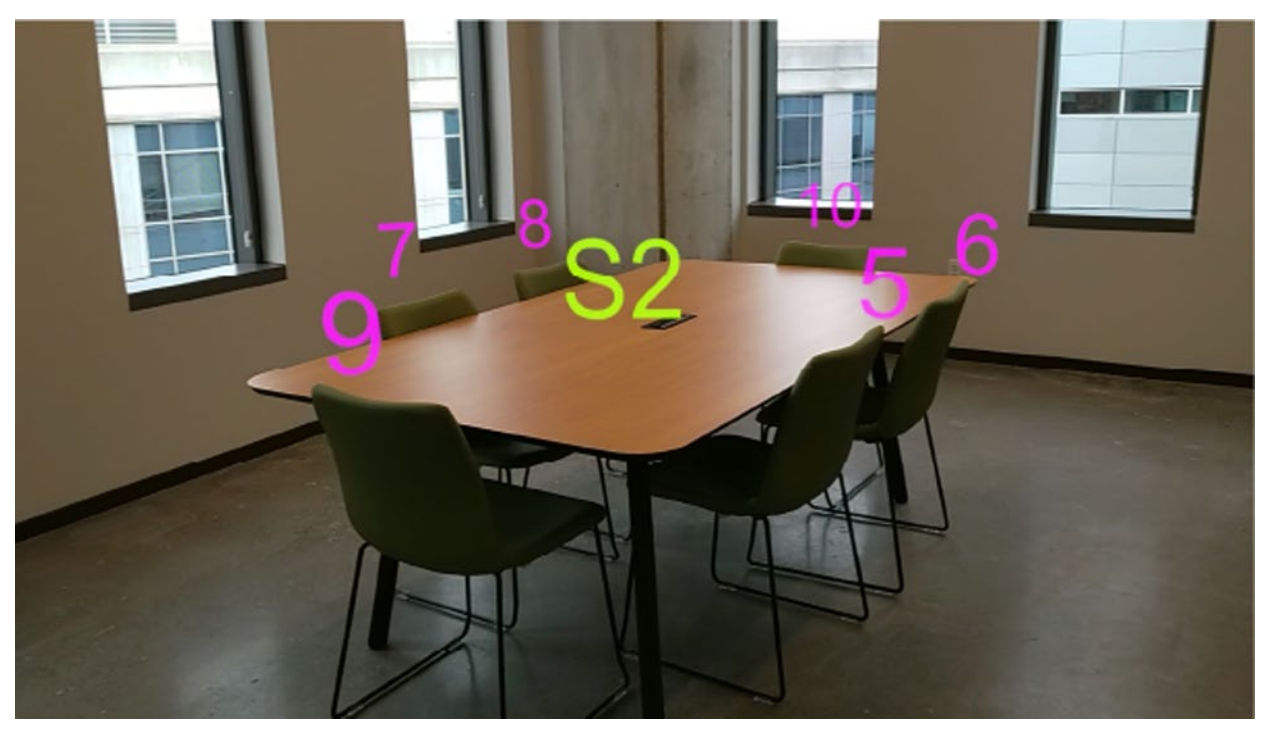

**Fig. 10** Layout visualization on HL2

Following the same procedure used in the synthetic experiments, the translation cost weight $w_\mathrm{t}$ and the rotation cost weight $w_\mathrm{r}$ were set to $0.36\sqrt{\mathrm{scene_size}}$ and 1, respectively.

*Result Analysis* We extracted and analyzed the log files for each scene on the HoloLens 2. Table 5 reports the mean and standard deviation of the accuracy and time cost.

For the changed layout of the pantry scene (Fig. 13), the average accuracy was $80.82\% \pm 22.26\%$, with a

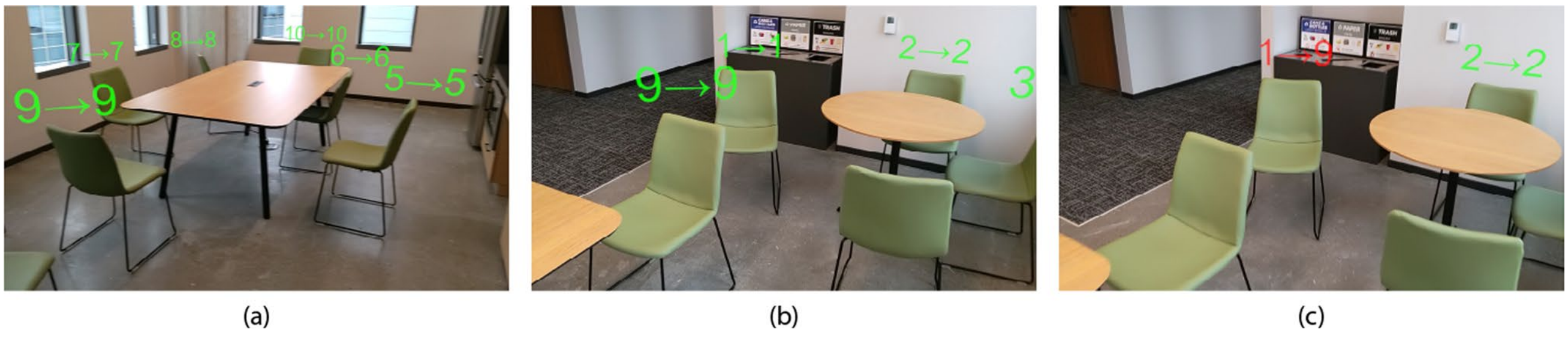

**Fig. 13** Example frames from the pantry scene. Please refer to the main text for detailed explanations

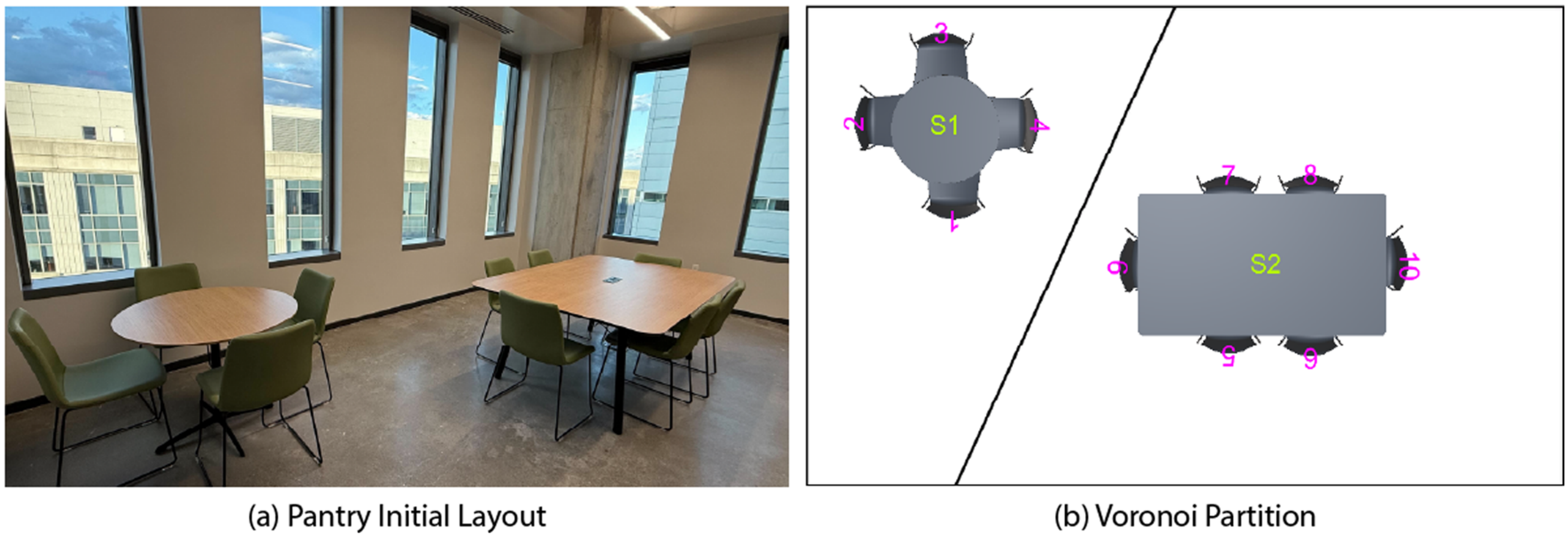

**Fig. 11** Initial layout of the pantry scene and its corresponding Voronoi partition illustration

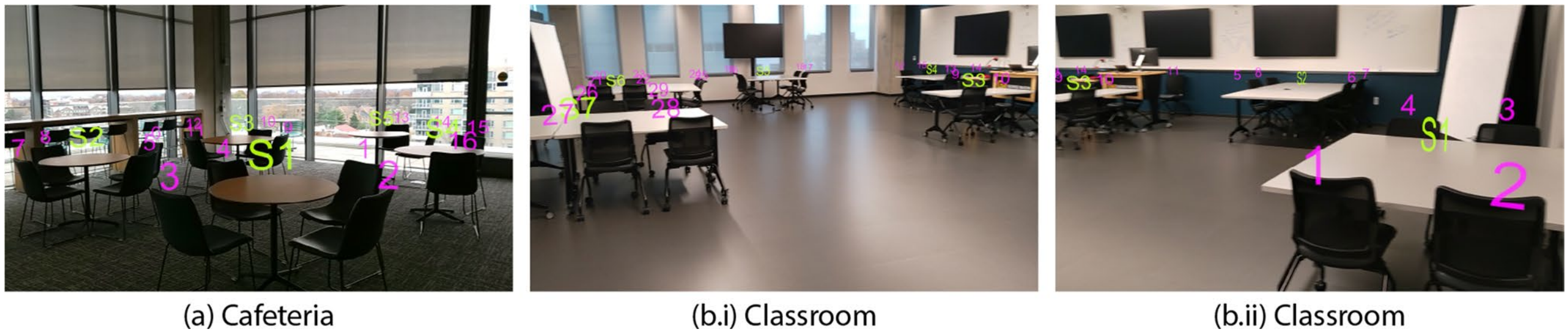

**Fig. 12** Voronoi partition visualizations for the initial layouts of the cafeteria and classroom scenes. Green labels (e.g., $S1$) indicate the Voronoi sites, while pink labels (e.g., 2) denote the assigned IDs in the initial layout. (a) The cafeteria scene. (b.i-b.ii) Two views of the classroom scene

**Table 5** Accuracy (%) and time cost (ms) for the three scenes. Accuracy refers to the ratio of correctly identified IDs to the total number of detected objects within a single frame, as defined in Eq. (10)

| Scene | Accuracy | | Time Cost | |
|---|---|---|---|---|
| | Mean | SD | Mean | SD |
| Pantry | 80.82 | 22.26 | 2.82 | 3.04 |
| Cafeteria | 88.51 | 18.05 | 4.43 | 3.73 |
| Classroom | 80.10 | 14.16 | 5.66 | 4.19 |

Time cost denotes the duration required to solve the IDs using integer programming on the HoloLens 2.

time cost of $2.82\,\text{ms} \pm 3.04\,\text{ms}$. For the cafeteria scene (Fig. 24), the accuracy was $88.51\% \pm 18.05\%$, and the time cost was $4.43\,\text{ms} \pm 3.73\,\text{ms}$. For the classroom scene (Fig. 25), the corresponding accuracy and time cost were $80.10\% \pm 14.16\%$ and $5.66\,\text{ms} \pm 4.19\,\text{ms}$, respectively.

Following the synthetic experiments, the accuracy is influenced by the magnitude of change between the initial and the changed layouts. In general, greater changes lead to lower accuracy. Regarding the standard deviation in accuracy, the number of detected objects (i.e., those within the FOV) plays a significant role. For example, consider two cases: (a) if 9 out of 10 detected objects are correctly identified, the accuracy is $90\%$; (b) if only 1 out of 2 detected objects is correctly identified, the accuracy is $50\%$. Although both cases involve a single incorrectly assigned ID, the resulting accuracies differ substantially due to the varying total counts.

Example frames for the pantry scene are shown in Fig. 13, which also depicts the changed layout. Labels follow the format $a \rightarrow b$, where $a$ is the assigned ID from the initial layout and $b$ is the solved ID in the changed layout. Green labels indicate correct assignments, while red labels indicate incorrect ones. Chairs without floating labels are not detected by our approach, as their bounding boxes are not fully within the FOV. In Fig. 13b, the label for chair 3 appears truncated due to the camera's clipping plane.

A comparison between Fig. 13b and c demonstrates that increased observations can lead to improved results. Chair #1 is correctly assigned in Fig. 13b (green $1 \rightarrow 1$), but incorrectly assigned in Fig. 13c (red $1 \rightarrow 9$). This outcome is reasonable from the perspective of integer programming, as the cost for assigning $1 \rightarrow 1$ is higher than that for $1 \rightarrow 9$ in frame (c).

Additional example frames for the cafeteria and classroom scenes are provided in Appendix B.

The accuracy distributions for the three scenes are shown in Fig. 14. As observed, accuracies of $100\%$ account for the largest proportion across all three scenes. Differences in the distributions arise from several factors, including scene size, the number of movable objects (i.e., chairs in our setup), the initial layout, and the magnitude of changes applied. Under the given real-world experimental settings, our approach consistently achieves an overall accuracy exceeding $80\%$.

## 4.3 Qualitative real-world experiments

We developed a prototype Farm-to-Table AR game utilizing the Microsoft HoloLens 2 headset to validate our proposed approach. Additionally, we presented two potential applications that could benefit from our approach. One application pertains to AR storytelling, while the other pertains to a delivery robot. Our experiments demonstrate that our approach is not limited to tabletop AR games. It is also well-suited for room-scale AR games, particularly in scenes featuring identical objects.

Furthermore, our approach shows potential in robot delivery, especially when the task involves identifying identical objects.

### 4.3.1 Farm-to-table AR game

In reference to the mobile farm-to-table game, Egg Inc AR Experience (Fig. 15), we developed an AR headset-based Farm-to-Table AR game to showcase our approach. The game incorporates various elements, including environmental animals (goats and cattle), controlled animals (chickens and bunnies), environmental props (silos, troughs, grasses, wood stacks, and flowers), and buildings (chicken coops, bunny hutches, and an animal shed).

*Game Settings* Fig. 16 illustrates the game settings where we manually divide the region into 4 Voronoi sites, namely S1 to S4. The player, wearing an AR headset (Microsoft HoloLens 2), is positioned in S3. The animal shed serves as the starting point for chickens and bunnies. Chickens move towards chicken coops while bunnies head towards bunny hutches. Both chickens and bunnies come in 5 different colors, and the corresponding coops and hutches also have matching colors. Each colored chicken or bunny moves toward the coop or hutch of the same color. Each coop or hutch has a label floating above it in the format of *animal building (count)*. Here, *animal building* indicates the type of colored animal attracted to the building, and *count* represents the number of such colored animals present. The *count* increases by one each time an animal with the corresponding color enters.

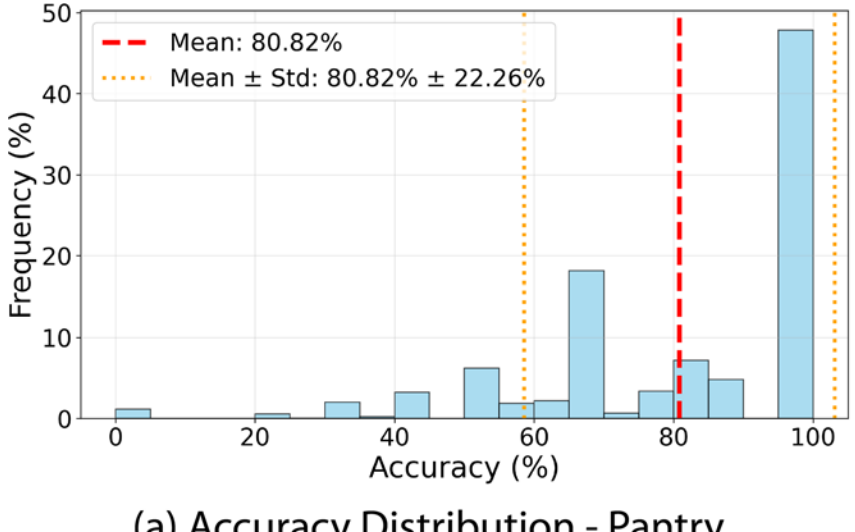

(a) Accuracy Distribution - Pantry

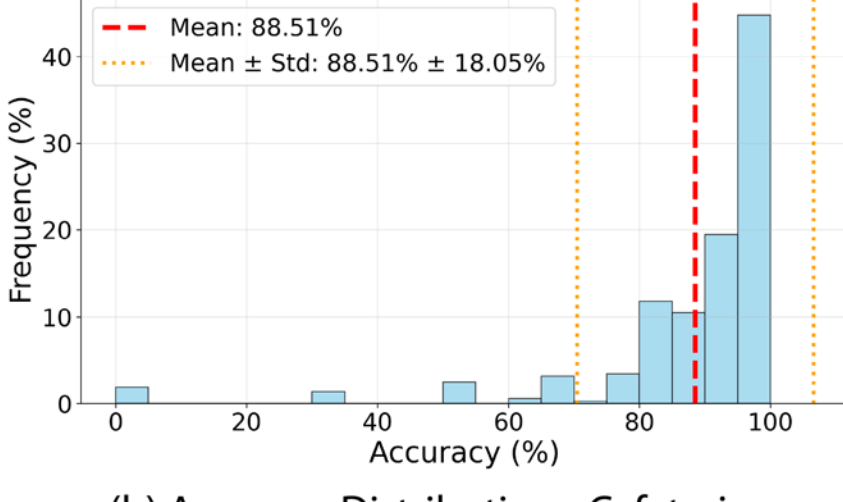

(b) Accuracy Distribution - Cafeteria

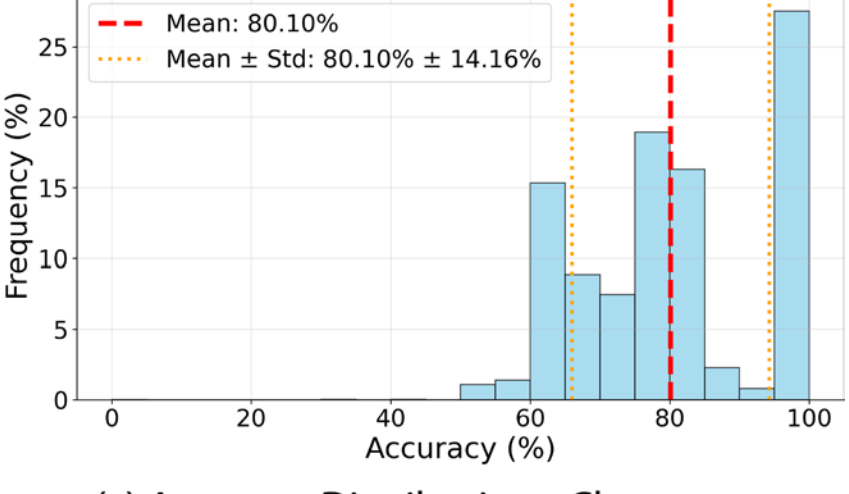

(c) Accuracy Distribution - Classroom

**Fig. 14** Distributions of accuracy for the three scenes. Please refer to the main text for further details

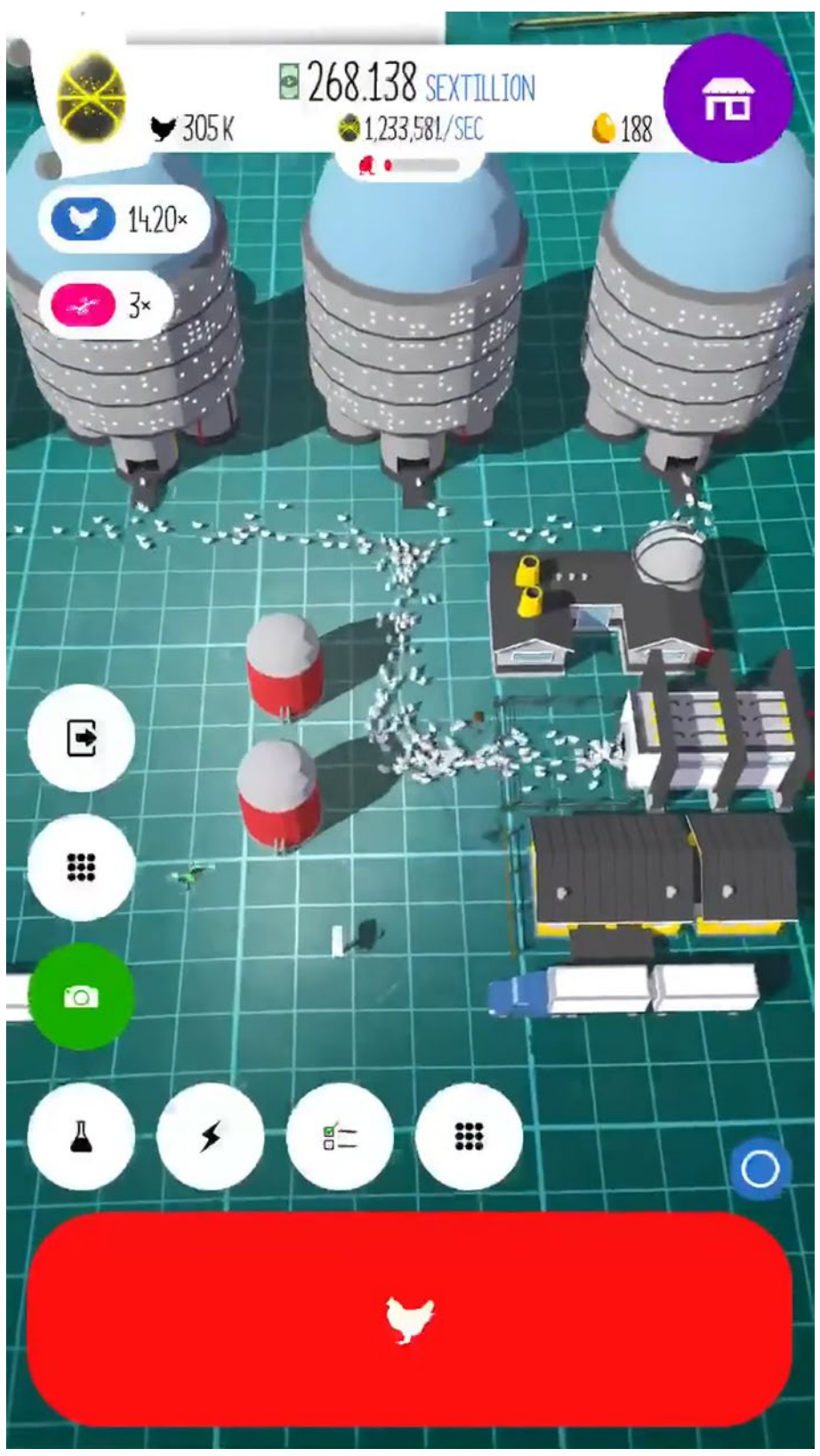

**Fig. 15** A screenshot of the Egg Inc AR Experience (Original gameplay video: https://www.youtube.com/shorts/3S6_ZNuXzU4)

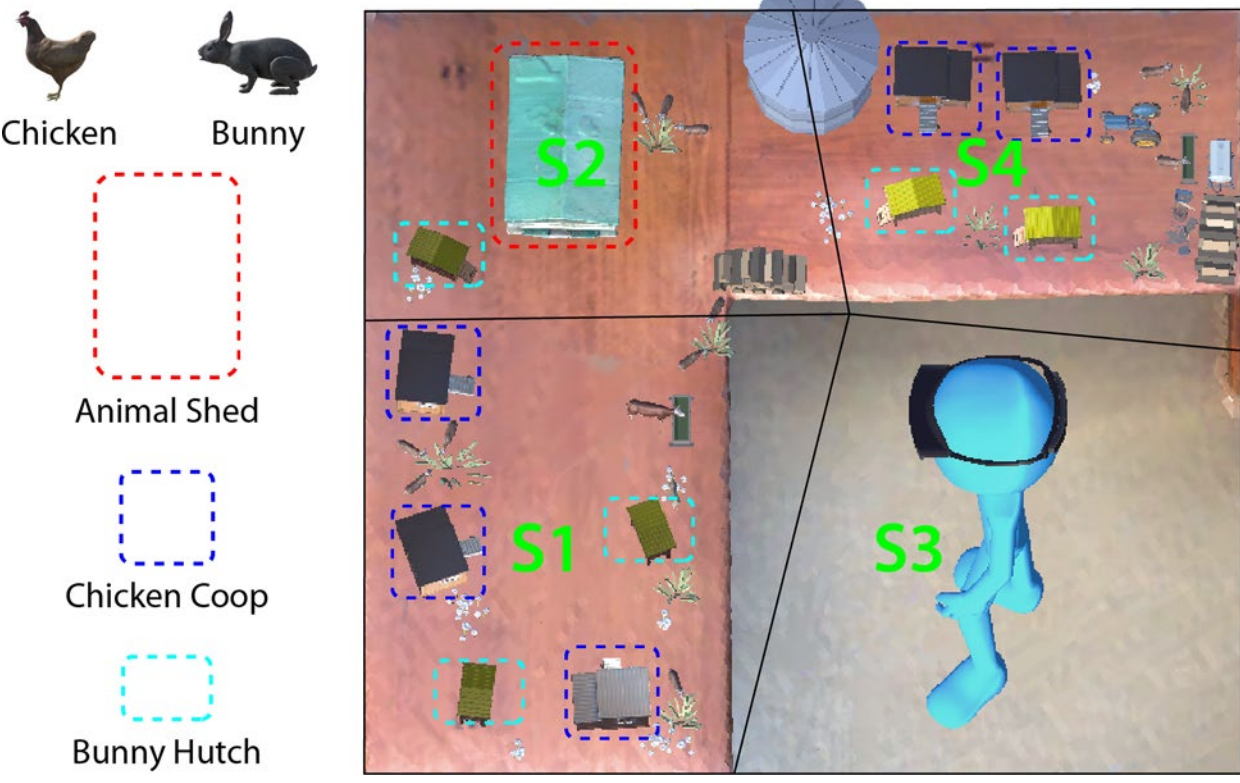

**Fig. 16** Game settings of our illustrative Farm-to-Table AR game

In our experiment, only the chicken coops, bunny hutches, and the animal shed physically exist.The player can reposition chicken coops and bunny hutches, but the animal shed remains fixed in its location. This restriction is due to the singular identity of the animal shed, which remains constant regardless of the layout changes.

Note that while we initialize a Voronoi diagram, we set the pruning threshold to 1.0. This choice implies that all sites will be considered for integer programming. This decision is deemed reasonable given the relatively small size of the scene, where the trade-off between time and accuracy is deemed unnecessary. Regarding the resolution of the label assignment problem, we address coops and hutches separately since a recognized coop cannot be a hutch and vice versa. In this process, we set the weight ratio from translation cost $w_t$ to rotation cost $w_r$ at a 1:1 ratio.

In this game setting, we follow the typical AR game procedure by initially scanning and mapping the table layout. Subsequently, the game begins, allowing the player to reposition the physical coops and hutches. The animation of the chickens and bunnies freezes when the player's hand becomes visible in the camera view. After the player completes the repositioning and his hand is no longer in view, the headset camera captures an image and records the camera pose. This pose is later used for mapping the camera pose when capturing the image. The poses of coops and hutches in the captured image are then resolved. Following this, our object label assignment algorithm identifies their respective identities and updates the game accordingly.

*Gameplay Demonstration #1* In this demonstration, the player is positioned facing the Vonoroi site S1 as depicted in Fig. 16. Figure 17 provides a visual representation of this demonstration. The virtual environment includes objects such as environmental props (e.g., grasses, flowers, troughs, wood stacks), environmental animals (e.g., goats grazing on grass, a wandering dog, and a cattle drinking water from the trough), controlled animals with various colors (chickens and bunnies), and text labels with diverse colors. On the other hand, real objects are also present, featuring an animal shed, three chicken coops, and three bunny hutches. Chickens and bunnies are observable in the scene, moving toward their respective coops and hutches. Figure 17a and b depict the initial game scene and the updated game scene, respectively.

Analyzing the output of our object label assignment algorithm, presented in Fig. 17b, we can infer the following layout changes: (1) swapping the positions of the *Blue Chicken Coop* and the *Cyan Bunny Hutch*; (2) relocating the *Magenta Bunny Hutch* to the left of the *Cyan Bunny Hutch*; and (3) slightly pushing the *Yellow Chicken Coop* further

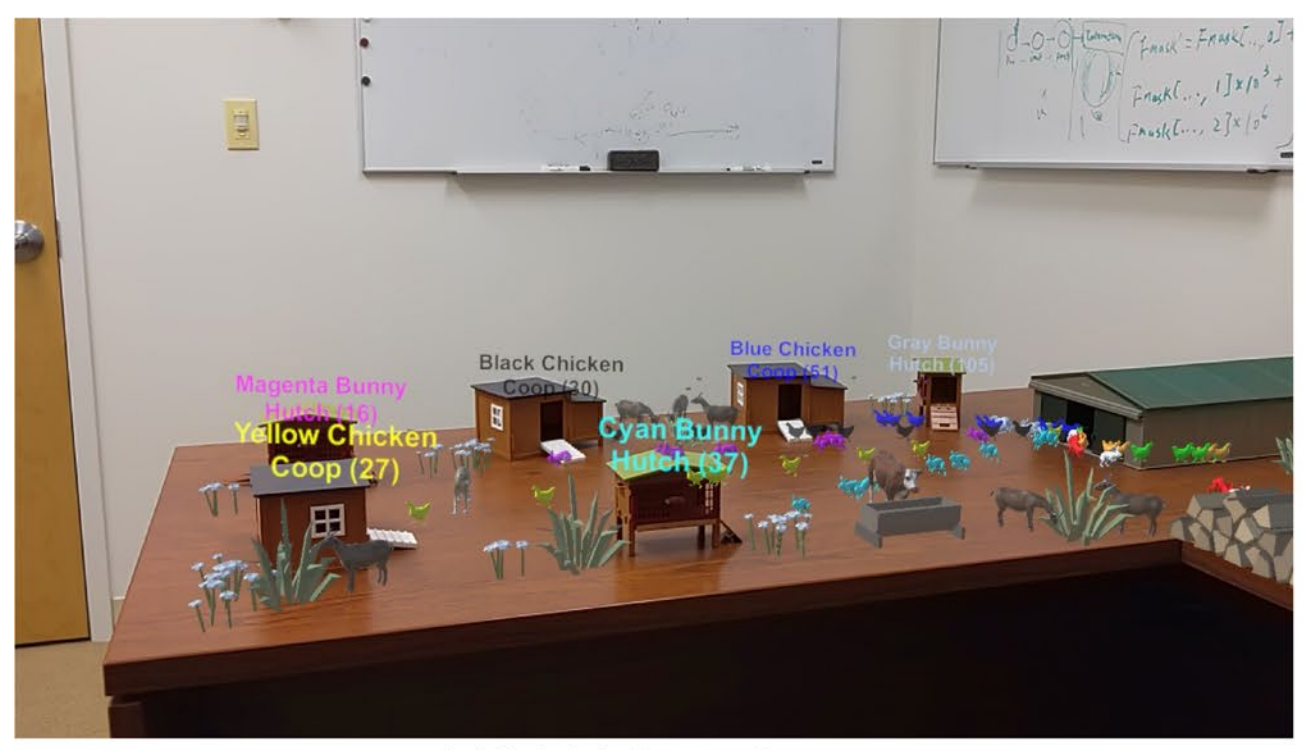

(a) Initial Game Scene

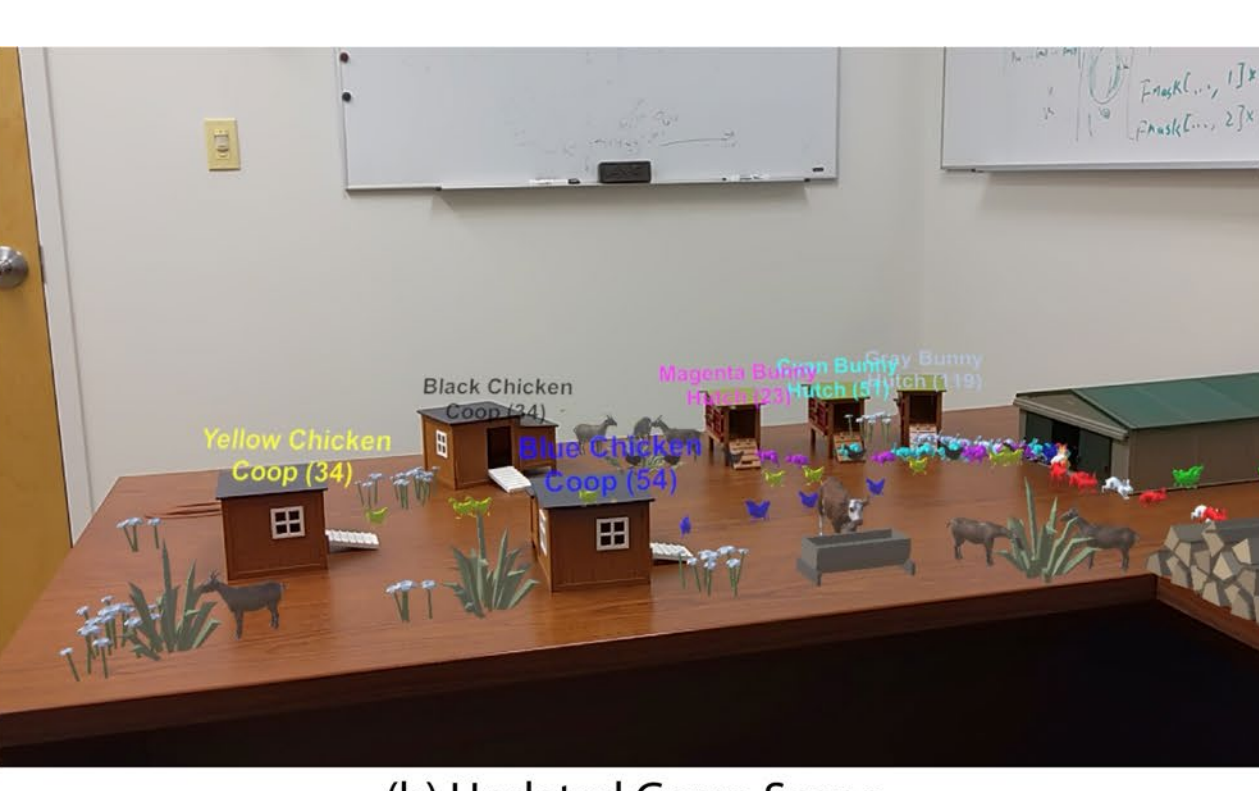

(b) Updated Game Scene

**Fig. 17** A demonstration of our AR headset Farm-to-Table game when the player faces Voronoi site S1 (Fig. 16). **a** illustrates the initial game scene before any layout change, while **b** portrays the updated game scene after executing the layout changes and running our object label assignment algorithm. We can deduce the following layout adjustments from (**b**): (1) swapping Blue Chicken Coop and Cyan Bunny Hutch positions; (2) moving Magenta Bunny Hutch left of Cyan Bunny Hutch; and (3) slightly shifting Yellow Chicken Coop away from the headset

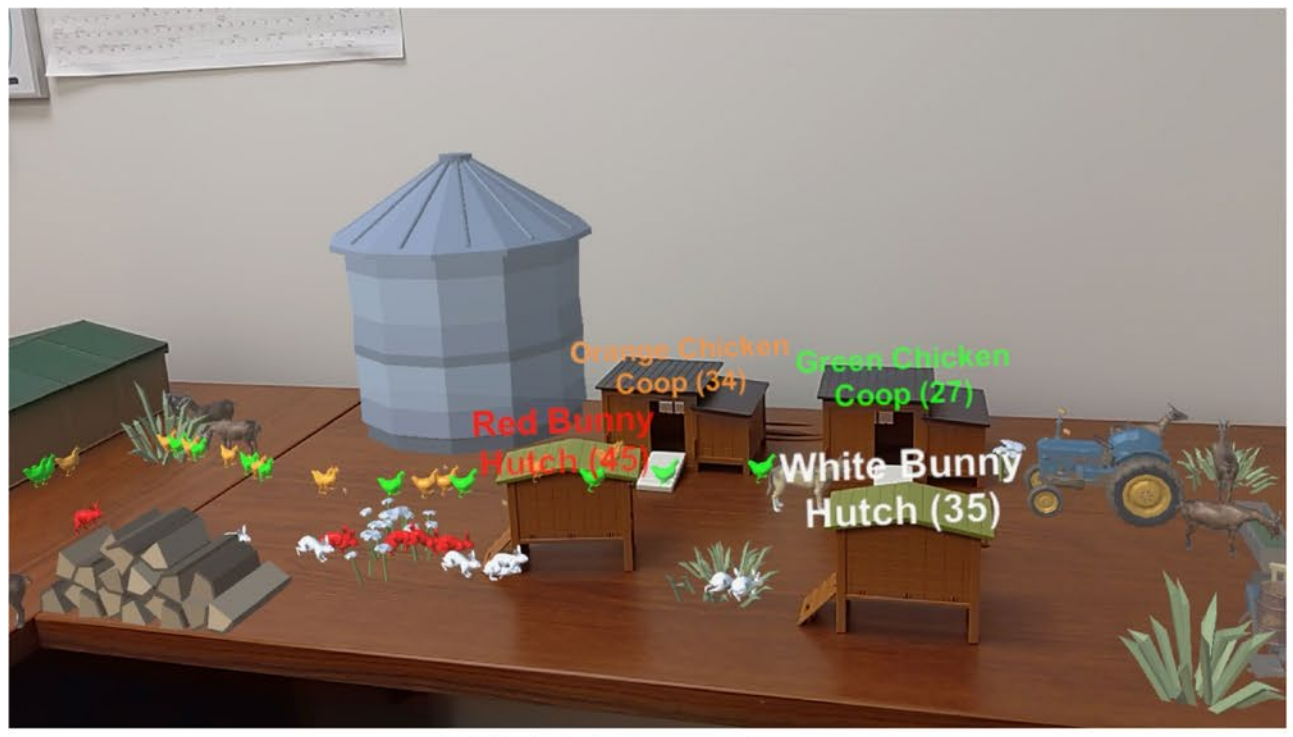

(a) Initial Game Scene

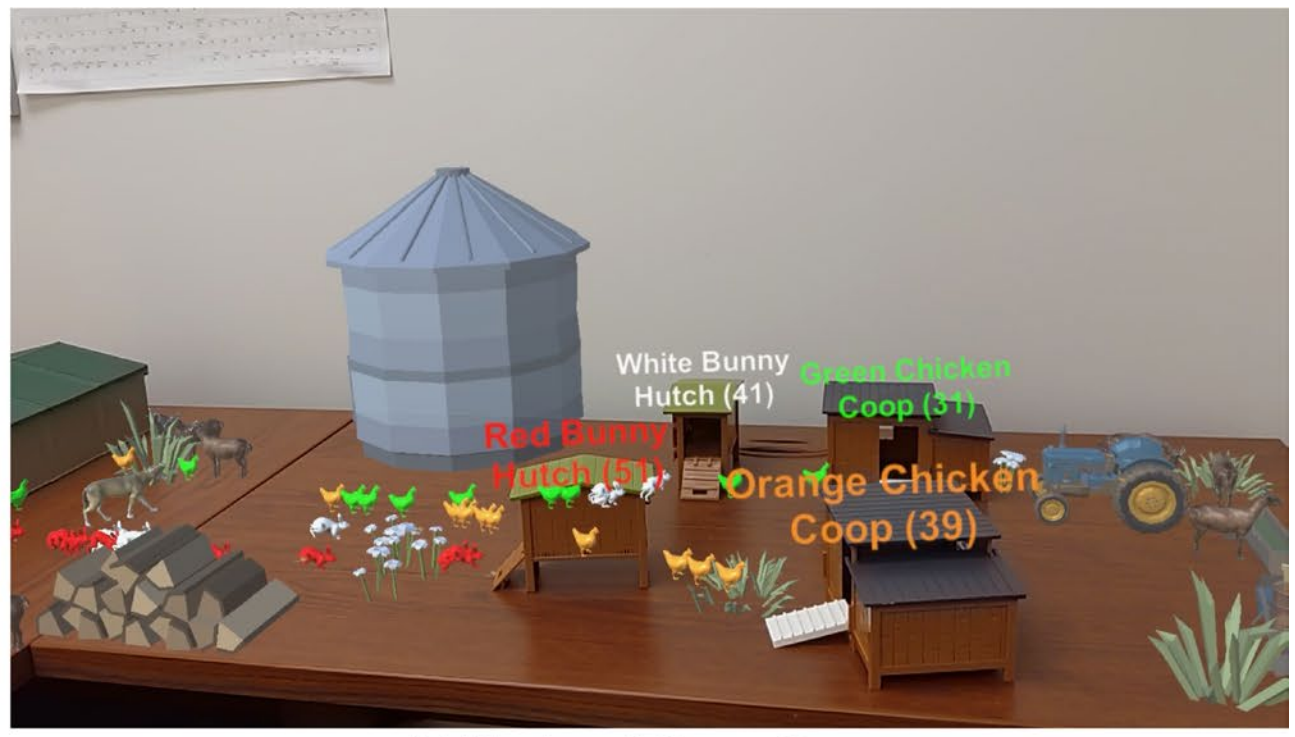

(b) Updated Game Scene

**Fig. 18** A demonstration of our AR headset Farm-to-Table game when the player faces Voronoi site S4 ( Fig. 16). **a** illustrates the initial game scene before any layout change, while **b** portrays the updated game scene after executing the layout changes and running our object label assignment algorithm. From the label assignment result, we can infer two layout changes: (1) swapping Orange Chicken Coop and White Bunny Hutch positions, and (2) rotating these objects

away from the headset. Considering our proposed cost terms, namely translation cost and rotation cost, the assignment result appears reasonable from a cost perspective.

*Gameplay Demonstration #2* In this demonstration, the player is facing the Voronoi site S4. Figure 18 provides a visual representation of this demonstration. The virtual environment includes objects such as environmental props (e.g., silo, grasses, flowers, troughs, wood stacks, tractors), environmental animals (e.g., goats grazing on grass, a wandering dog, a goat drinking water from the trough), controlled animals with various colors (chickens and bunnies), and text labels with diverse colors. On the other hand, real objects are visible, featuring part of the animal shed, two chicken coops, and two bunny hutches. Chickens and bunnies can be observed in the scene, which are moving towards their respective coops and hutches. Figure 18a and b depict the initial game scene and the updated game scene, respectively.

Examining the output of our object label assignment algorithm, presented in Fig. 18b, we can deduce the layout changes: (1) exchanging the positions of the *Orange Chicken Coop* and the *White Bunny Hutch*; and (2) rotating these objects. Evaluating these changes from the perspective of our proposed cost terms–translation cost and rotation cost–this assignment result is reasonable in terms of minimizing the associated costs.

#### 4.3.2 AR storytelling showcase

We use a partially scanned lab layout to showcase our approach's application for AR storytelling. First, we scanned and mapped the partial layout. We also modeled a 3D chair akin to the lab's real chairs. In this experiment, we moved the chairs in the lab. For pose estimation, we used *MediaPipe Objectron* (Ahmadyan et al. 2021).

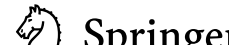

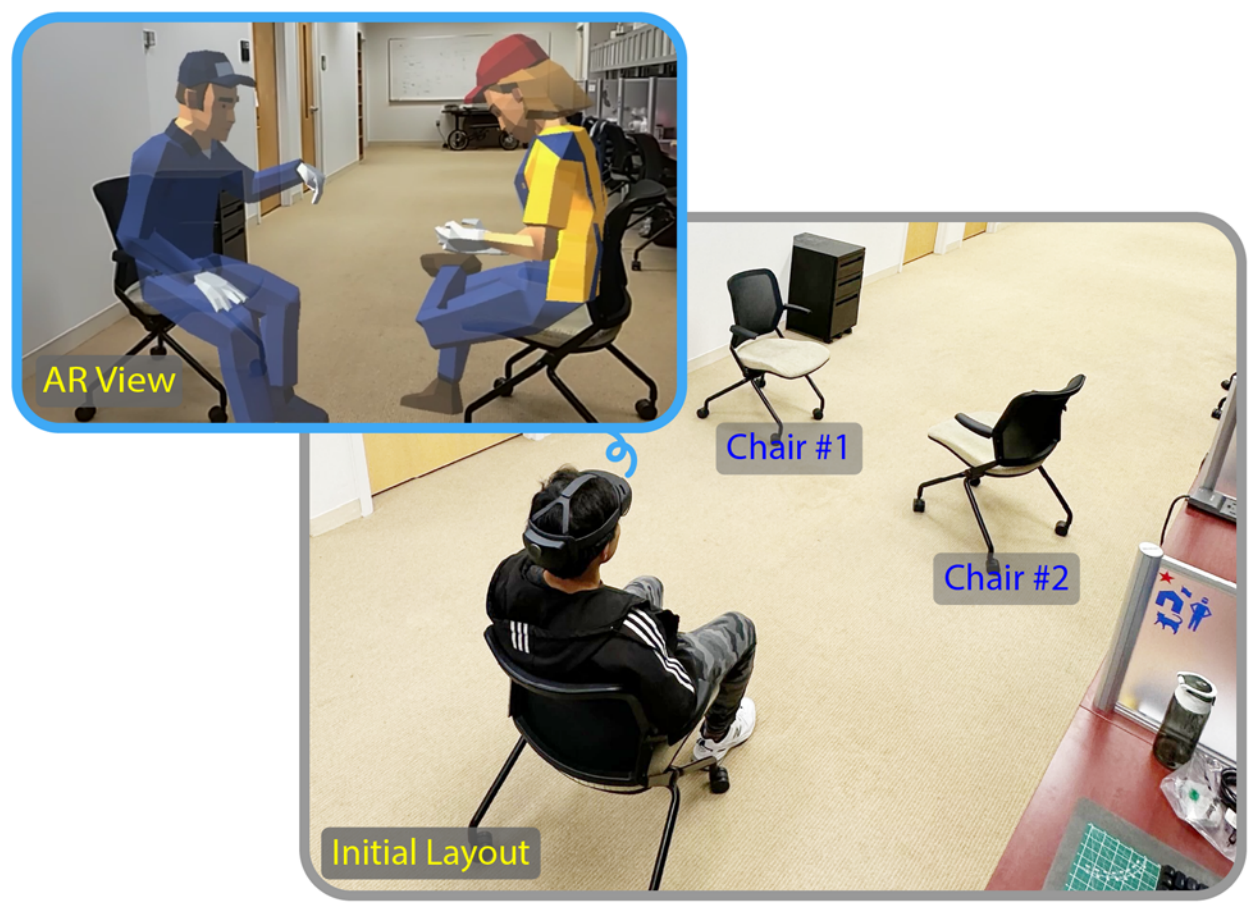

(a) Initial AR Storytelling Scene

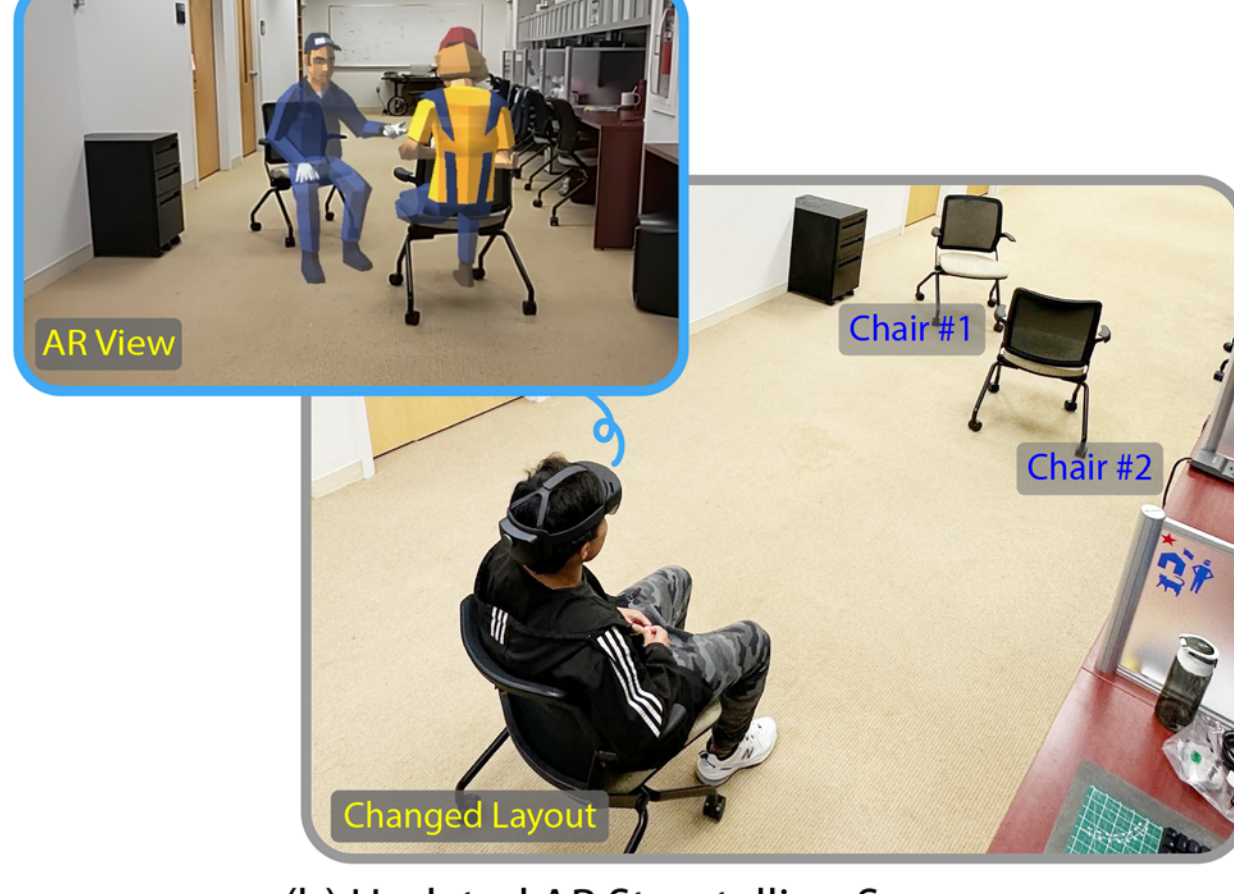

(b) Updated AR Storytelling Scene

**Fig. 19** An AR storytelling example. **a** The initial view of an AR story where two virtual characters sat on real chairs. **b** An updated AR view showing that the two virtual characters still sat on the same chairs as before in the changed layout

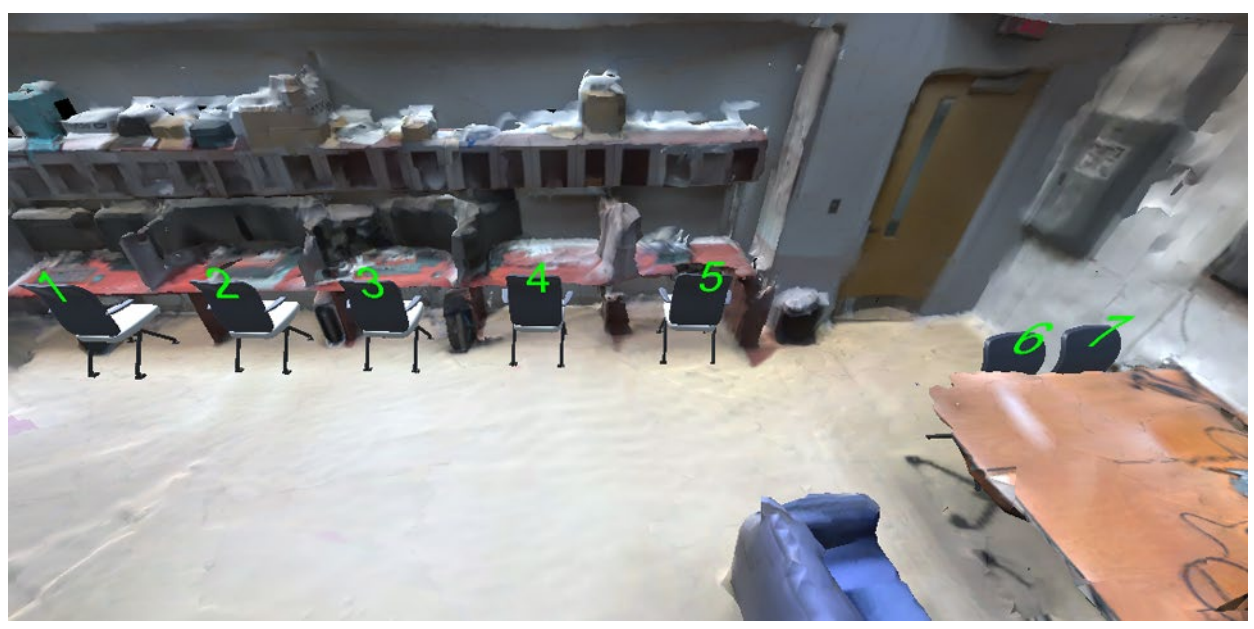

**Fig. 20** The initial layout of the lab environment. The numbers from 1 to 7 indicate movable chairs

We implemented an AR storytelling scenario as shown in Fig. 19. A user was watching an AR story where two virtual characters sat on two real chairs (Fig. 19a). More specifically, the male character sat on Chair 1, and the female character sat on Chair 2. After some time, Chair 1 and Chair 2 were manipulated by a person passing by. It would look unnatural if the program did not adjust the two virtual characters, as they would then be sitting in the air, not on the real chairs.

The AR headset that the person used for watching the AR story captured an image of the scene. The pose estimation algorithm took this image as input and output 6DOF poses. Our approach utilized the AR headset's camera pose and the pose estimation results to determine the chairs' identities, based on which the two virtual characters' positions and rotations (Fig. 19b) were updated. As a result, the two virtual characters still appeared sitting on the same chairs as before.

#### 4.3.3 Robot showcase

Based on the initial layout shown in Fig. 20, we simulated a delivery robot. The objective for the robot is to make a delivery to a designated chair. We used LoCoBot[2] as the robot platform. It was equipped with an Intel Realsense Depth Camera D435, which captured images for pose estimation. We set Chair 4 as the target that the LocoBot should move to.

For the experiment procedure, we first set up the robot, gave it a delivery target, and let it move forward. As our approach detected the target after pose estimation and label assignment, a target position was issued to the robot. Because of that, the robot would cancel the current goal first and execute the new goal, i.e., moving toward the target.

Figure 21a shows a changed layout. Figure 21b shows the object IDs inferred by our approach. The robot reached the expected target, showing that our approach performed as expected in the real environment.

## 5 Discussion

Our approach distinguishes visually identical objects using an egocentric partial observation frame captured by an AR headset. It can also incorporate a pruning algorithm to process a large environment with numerous objects more efficiently. Its effectiveness has been validated across a spectrum of simulated environments, spanning various levels of complexity, and an ablation experiment. We implemented a Farm-to-Table game based on an AR headset, demonstrating the practical application of our approach. Moreover,

[2] LoCoBot: http://www.locobot.org/

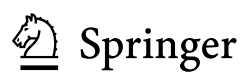

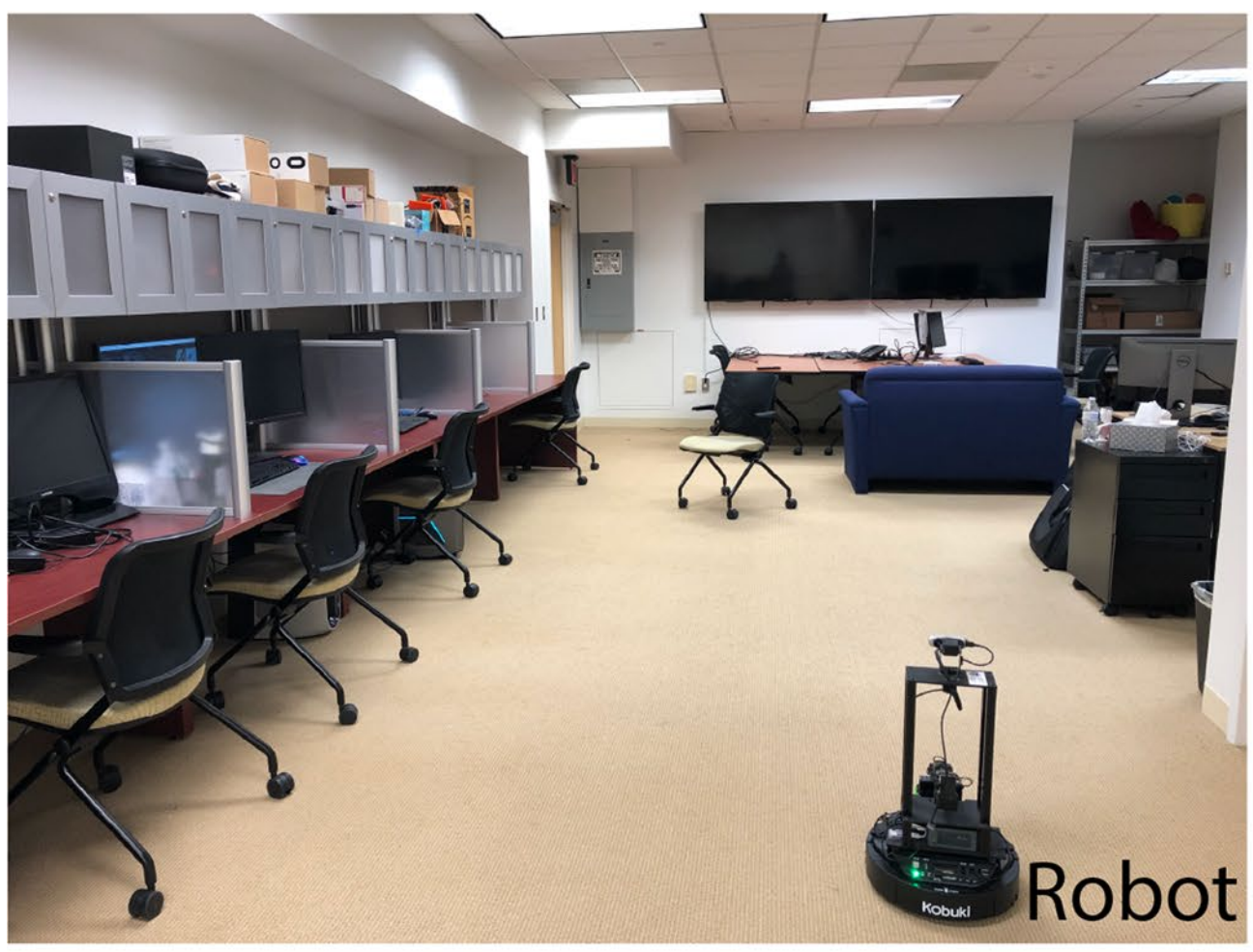

(a) Changed Layout

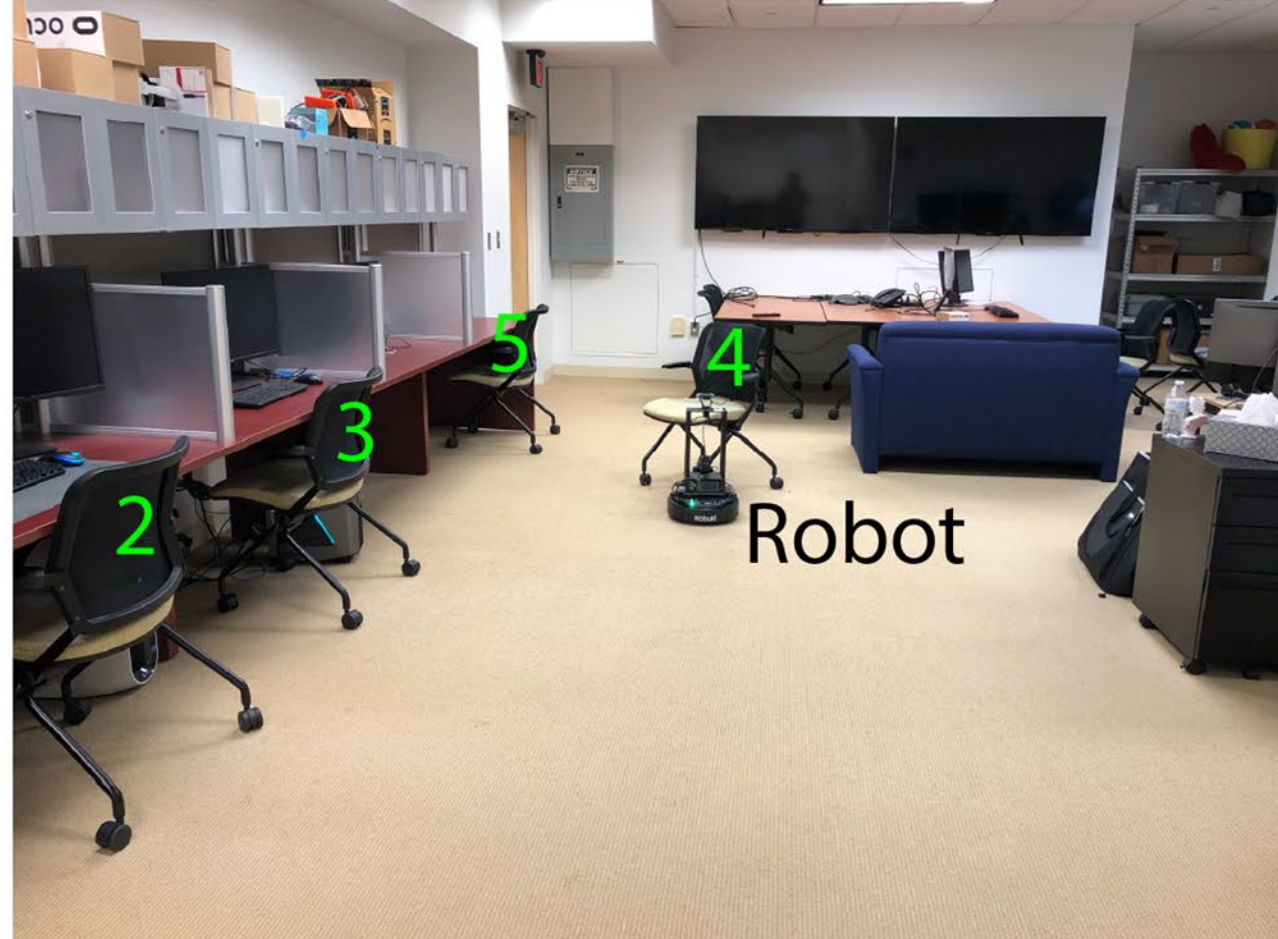

(b) Robot Reaching Target

**Fig. 21** The robot simulation. **a** The changed layout in the real world at the beginning; **b** the scene when the robot reaches the target in the end

our approach can smoothen AR storytelling experiences by leveraging tracked real objects. We also demonstrated our approach for a simulated robot delivery scenario.

## 5.1 Limitations

Within the realm of AR games, user experience holds paramount importance, with immersiveness serving as a key contributing factor. While our approach proficiently resolves the identities of identical objects following layout changes, it does not generate the dynamic AR visual effects introduced by such changes, such as light variations and shadows. Beyond object identity resolution, it is useful to extend the AR gaming pipeline to incorporate real-time visual updates based on the tracked objects during gameplay, which would contribute to a more compelling and immersive AR gameplay experience.

Our approach hinges on the precision of pose estimation. Any inaccuracy in this aspect can lead to mislabeled results. At the hardware level, the HoloLens 2 has a limited field of view, restricting its ability to capture multiple objects simultaneously, especially in large scenes. The more objects it captures, the more constraints are introduced, resulting in better integer programming outcomes in general. While our algorithm can handle the simultaneous assignment of multiple object identities, its effectiveness depends on the number of objects that can be detected. Furthermore, in terms of the pose estimation algorithm, utilizing a versatile pose estimation approach that does not require scanned meshes and texture could broaden the applicability of our approach.

In this paper, our assumptions revolve around the movement and rotation capabilities of objects within our model. We consider objects capable of movement along two primary directions: the $x$-axis, corresponding to front-and-back motion, and the $z$-axis, corresponding to left-and-right motion. We do not account for movements in the up-and-down direction in our model. Furthermore, our assumption restricts the rotation of objects to occur solely about the $y$-axis. In some real-world scenarios, objects may have more degrees of freedom when it comes to movement and rotation. In reality, objects can rotate freely about various axes, leading to a more complex and versatile range of spatial transformations. Our simplified assumptions are made for the sake of modeling simplicity and computational efficiency, but should be understood in the context of these real-world limitations.

## 5.2 Future work

Currently, if the layout is changed by the player, a game pause is required to await the completion of layout changes, perform pose estimation, and subsequently update the game in our implementation. However, this pause in gameplay may impact user experience. To refine this aspect, a real-time update is preferable. This entails synchronizing the game updates with the ongoing changes in the real layout, eliminating the need for pauses. By adopting a real-time update strategy, it is possible to seamlessly integrate alterations to the physical environment with the virtual elements, enhancing the user experience by providing a more fluid and continuous gameplay interaction.

In our setup, real-world objects function as NPCs, with virtual objects interacting with them dynamically. NPCs often host events in games, and enhancing player immersion involves detecting their actions and updating the game scene accordingly. For instance, imagine a scenario that triggers virtual chickens and bunnies to emerge right after the player opens the gate of the real animal shed. Similarly, it could be designed to prompt designated chickens to alter their routine and relocate to another coop once the player removes the stairs from a coop, which will add depth to the AR gaming experience.

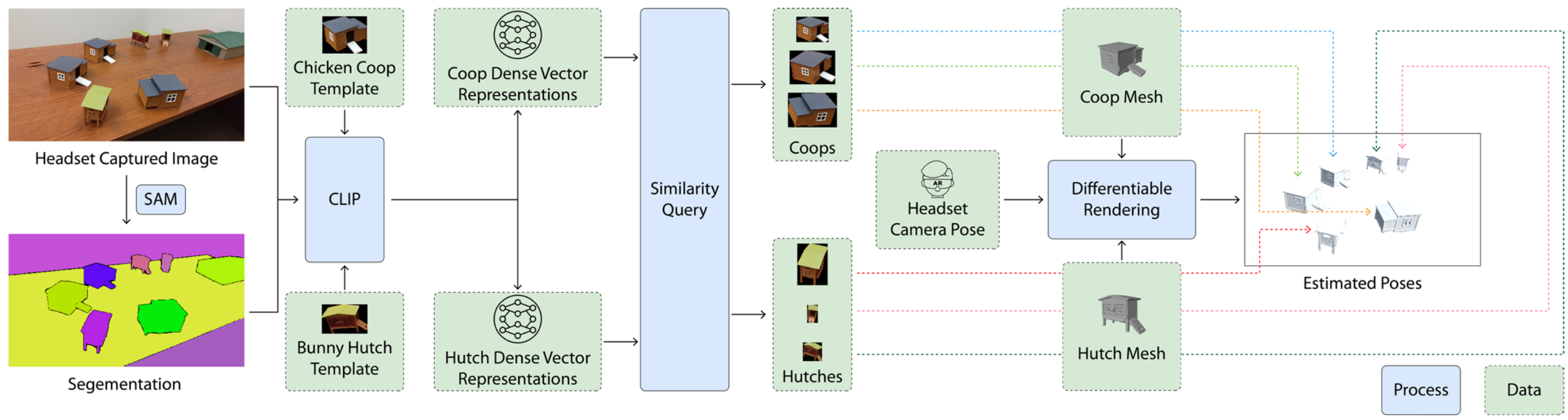

Fig. 22 The pipeline of pose estimation for our Farm-to-Table game. Please refer to the main text for the explanation

In our experiment, physical objects remain stationary during the gameplay. However, future AR games may involve scenarios where physical objects exhibit the same dynamism as virtual ones. Consider a cooking AR game where players manipulate real kitchen tools such as solid turners. The AR gaming experience would become more realistic and immersive if players could use authentic kitchen tools, such as turners and pans, to interact with virtual food elements. Such an experience would surpass the realism attained through hand-tracking or controller-based simulations.

## 6 Conclusion

We presented a novel optimization-based approach to enhance physical-virtual interaction in AR gaming by resolving the identities of visually identical real-world objects. Leveraging an egocentric partial observation frame captured by an AR headset, our method formulates the label assignment task as an integer programming problem, enabling identity tracking without continuous observation or external infrastructure. We validated its effectiveness through quantitative synthetic and real-world experiments, and demonstrated its practical utility and applicability via three qualitative real-world experiments: a farm-to-table AR game, an AR storytelling showcase, and a robot delivery showcase.

## Appendix A farm-to-table game pose estimation

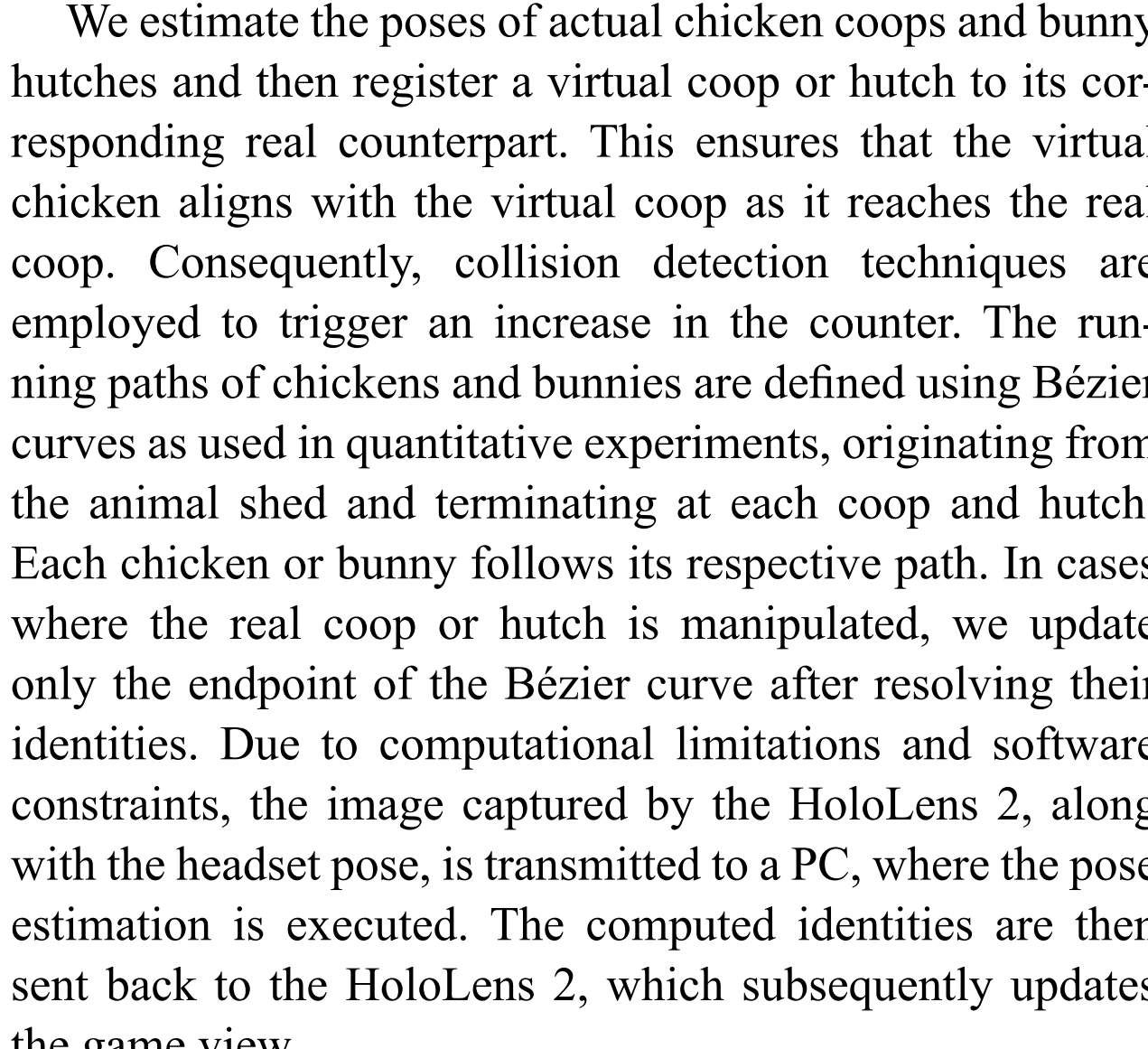

We estimate the poses of actual chicken coops and bunny hutches and then register a virtual coop or hutch to its corresponding real counterpart. This ensures that the virtual chicken aligns with the virtual coop as it reaches the real coop. Consequently, collision detection techniques are employed to trigger an increase in the counter. The running paths of chickens and bunnies are defined using Bézier curves as used in quantitative experiments, originating from the animal shed and terminating at each coop and hutch. Each chicken or bunny follows its respective path. In cases where the real coop or hutch is manipulated, we update only the endpoint of the Bézier curve after resolving their identities. Due to computational limitations and software constraints, the image captured by the HoloLens 2, along with the headset pose, is transmitted to a PC, where the pose estimation is executed. The computed identities are then sent back to the HoloLens 2, which subsequently updates the game view.

Given the absence of an off-the-shelf solution for estimating the poses of coops and hutches in our experiment, we devised a pipeline using state-of-the-art computer vision techniques. Figure 22 illustrates the pipeline. Initially, we pre-process the coops and hutches to obtain their meshes and partial textures. Using the captured image, we employ the SAM (Segment Anything Model) (Kirillov et al. 2023) to obtain segmentations. CLIP (Contrastive Language-Image Pre-Training) (Radford et al. 2021) is then utilized to calculate dense vector representations for each segmentation region, using the pre-trained *clip-ViT-B-32* model. The same process is applied to the partial textures (templates) of the coop and hutch (Fig. 23).

Subsequently, a similarity query approach is employed to identify regions akin to the templates. Once the coops and hutches are located, their corresponding binary masks in segmentation are used for the next step. Differentiable rendering, bridging the gap between 2D and 3D by linking 2D image pixels to the 3D properties of a scene (Keselman and Hebert 2022), is applied to solve the poses of coops based on the coop binary mask, the headset camera pose during image capture, and the coop mesh. The same procedure is followed for hutch masks. We used PyTorch3D (Ravi et al. 2020) for differentiable rendering. Once all poses are determined, our object label assignment algorithm runs to identify their respective object identities.

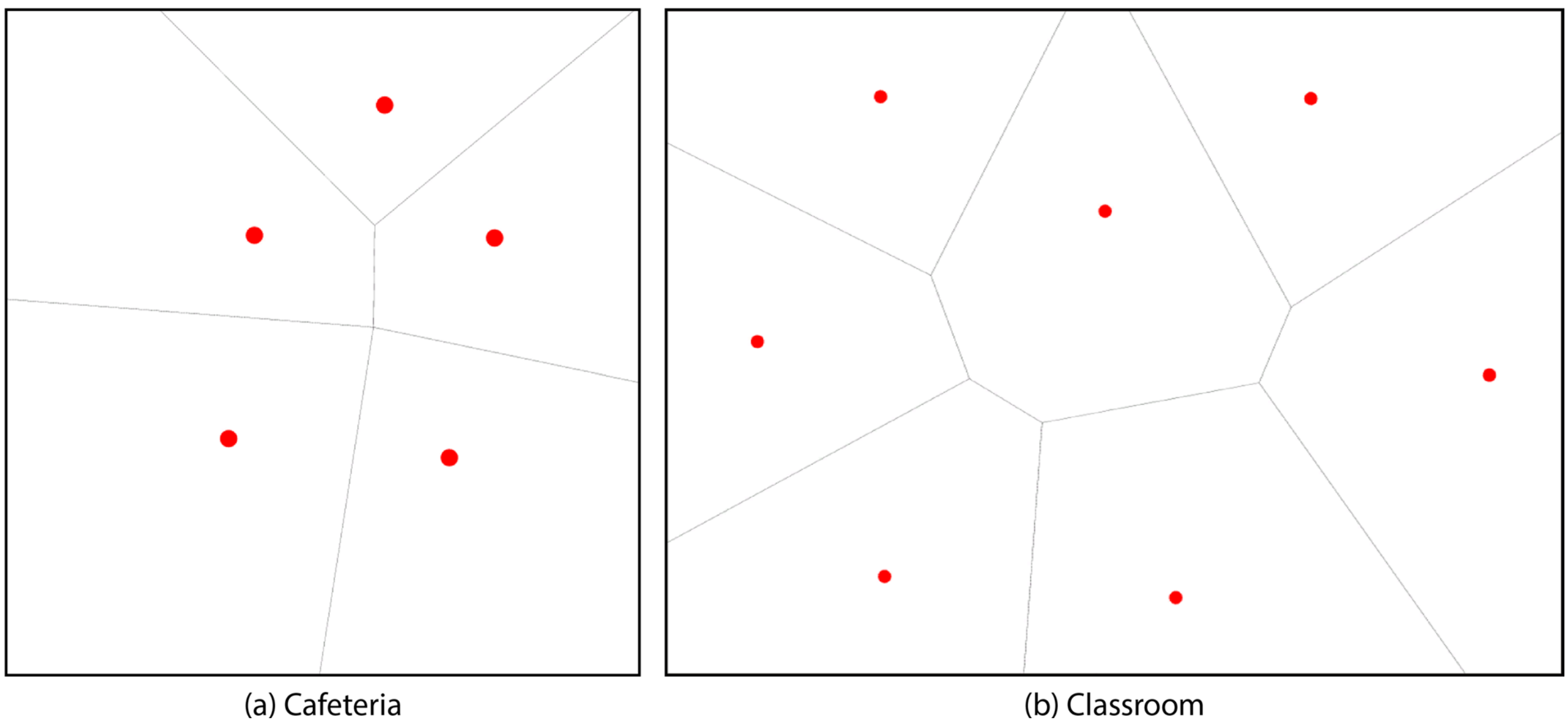

**Fig. 23** Voronoi diagram visualizations for the cafeteria and classroom scenes. Red dots represent the Voronoi sites (i.e., tables). Note that the images have been scaled for improved visualization

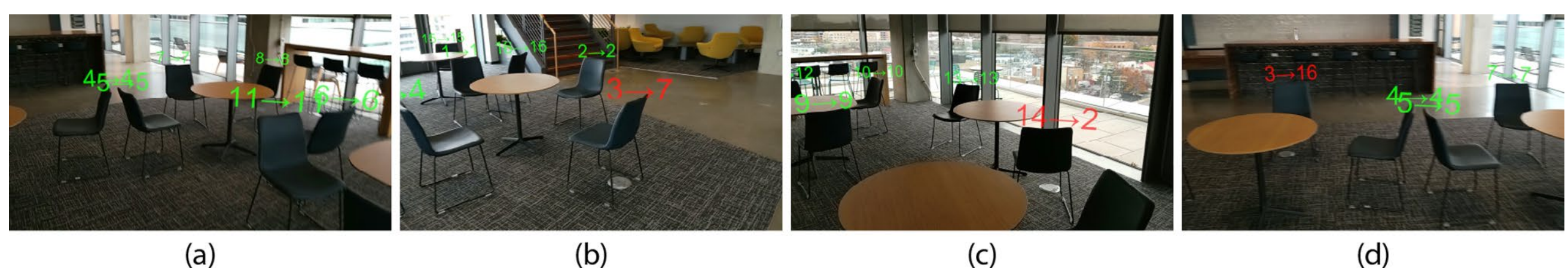

**Fig. 24** Example frames for the cafeteria scene

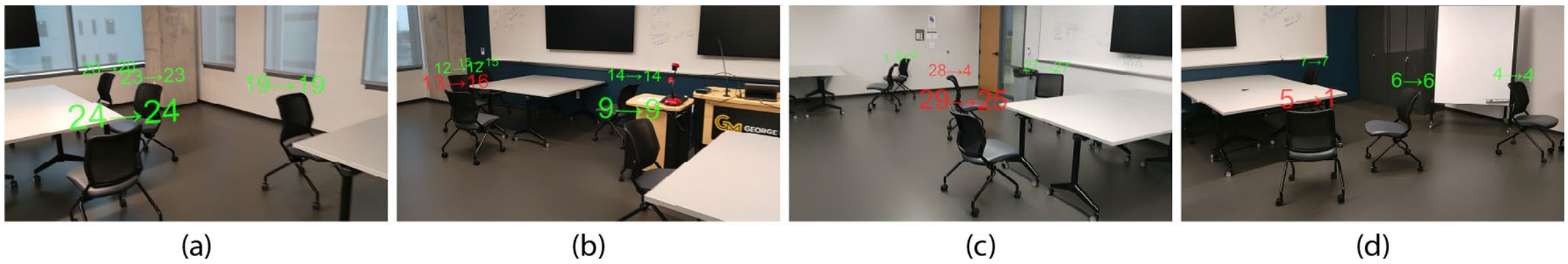

**Fig. 25** Example frames for the classroom scene

## Appendix B Cafeteria and classroom experiments

Voronoi diagram visualizations for the cafeteria and classroom scenes are shown in Fig. 23. Example frames for the cafeteria and classroom scenes are shown in Figs. 24 and 25, respectively. The initial layout of the cafeteria is depicted in Fig. 12a, while the initial layout of the classroom is shown in Fig. 12b.

Each label follows the format $a \rightarrow b$, where $a$ is the assigned ID from the initial layout and $b$ is the solved ID in the changed layout. Green labels indicate correct assignments, whereas red labels indicate incorrect assignments.

## Appendix C Robot showcase pipeline

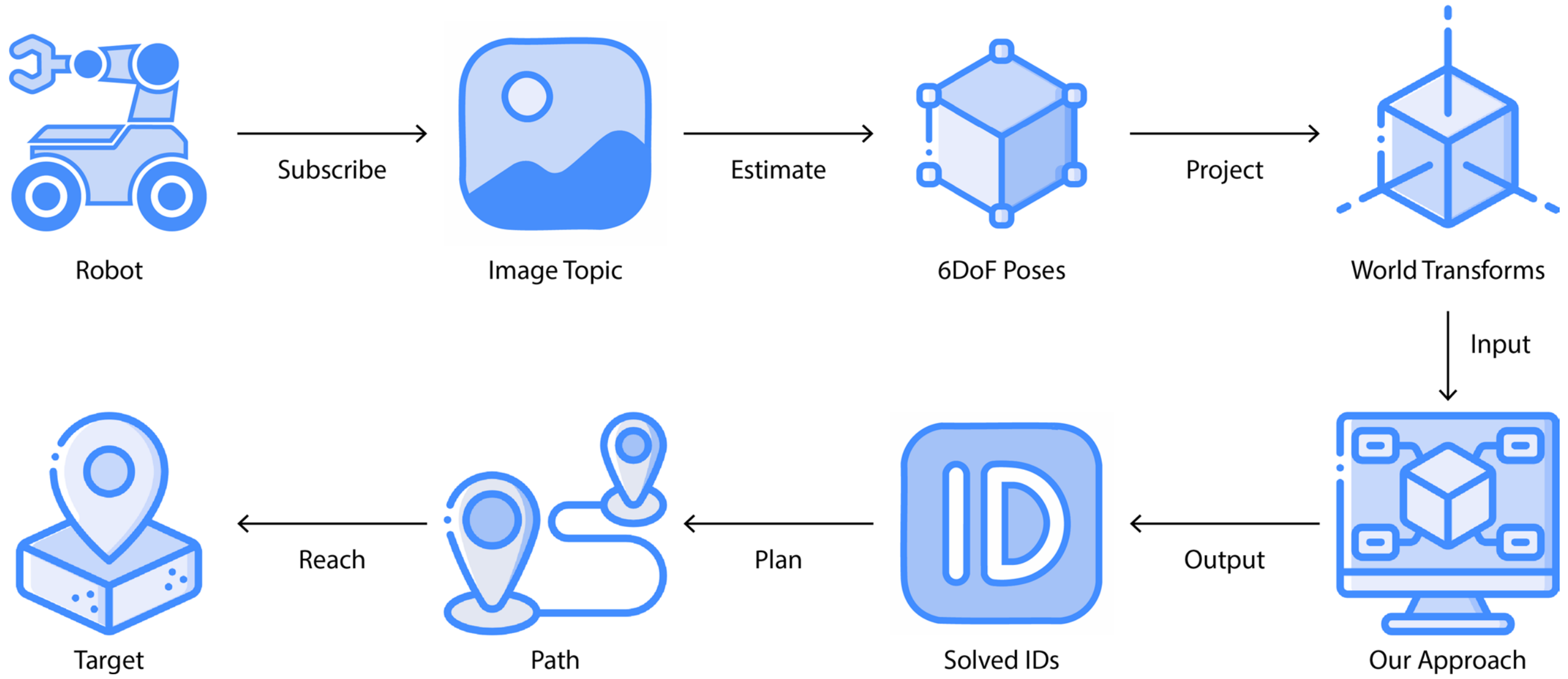

**Fig. 26** The pipeline of the robot showcase

Figure 26 illustrates the pipeline of the simulated robot delivery. The pose estimation was run on the RGB image. If it outputs a valid 6DOF pose, its corresponding depth image will be used to estimate the distance. After that, the world transformation of the object can be calculated using the camera's pose. Our object label assignment algorithm is then run to determine the object's identity label. If the target object is within the result, its position is sent to the robot, which then moves to the target.

**Supplementary Information** The online version contains supplementary material available at https://doi.org/10.1007/s10055-025-01305-y.

**Author contributions** L.Y.: writing, algorithm, experiments, and analysis; C.H.: robot showcase; H.W.: advising; Lap-Fai Y.: writing, advising, and project supervision.

**Funding** This project was supported by National Science Foundation grants with award numbers 1942531, 2418236, and 2430673.

**Data availability** Our approach is openly available in a GitHub Repository (Liuchuan Yu et al., https://github.com/gmudcxr/EnrichingARInteraction). The authors confirm that the data, tools, and documentation supporting the findings of this study are available within the article and through the links provided in the manuscript.

## Declarations

**Consent to participants** This work did not involve any human participants.

**Conflict of interest** The authors declare no conflict of competing for financial interests or personal relationships that could have appeared to influence the work reported in this paper